\documentclass[manuscript,authorversion,nonacm]{acmart}
\usepackage[inkscapeformat=png]{svg}
\usepackage{cleveref}
\usepackage{algorithm}
\usepackage{subcaption}
\usepackage{multicol}

\usepackage{tikz}
\usepackage{cleveref}
\usetikzlibrary{arrows}
\usepackage{circuitikz}
\usepackage{pdflscape}
\usepackage{framed} 
\usepackage{fancyvrb}
\usepackage{svg}
\usepackage{makecell}

\usepackage[colorinlistoftodos]{todonotes} 
\usepackage{booktabs} 
\usepackage{xmpincl}
\usepackage{subfiles}
\usepackage{algorithm}
\usepackage{algpseudocodex}

\AtBeginDocument{%
  }

\setcopyright{acmlicensed}
\copyrightyear{2018}
\acmYear{2018}
\acmDOI{XXXXXXX.XXXXXXX}
\acmConference[Conference acronym 'XX]{Make sure to enter the correct
  conference title from your rights confirmation email}{June 03--05,
  2018}{Woodstock, NY}
\acmISBN{978-1-4503-XXXX-X/2018/06}

\begin{document}

\title{Ray-Traced Augmentation for Signal Strength Based Localization}

\author{Jihoon Og}
\email{og@ualberta.ca}
\orcid{0009-0005-1824-9263}
\affiliation{%
  \institution{University of Alberta}
  \city{Edmonton}
  \state{Alberta}
  \country{Canada}
}

\author{Ningze Sun}
\email{ningze@ualberta.ca}
\orcid{0009-0004-9728-8756}
\affiliation{%
  \institution{University of Waterloo}
  \city{Waterloo}
  \state{Ontario}
  \country{Canada}
}

\author{Ioanis Nikolaidis}
\email{nikolaidis@ualberta.ca}
\orcid{0000-0003-1469-5280}
\affiliation{%
  \institution{University of Alberta}
  \city{Edmonton}
  \state{Alberta}
  \country{Canada}
}

\author{Omid Ardakanian}
\email{oardakan@ualberta.ca}
\orcid{0000-0002-6711-5502}
\affiliation{%
  \institution{University of Alberta}
  \city{Edmonton}
  \state{Alberta}
  \country{Canada}
}

\begin{abstract}
Indoor localization based on Wi-Fi typically relies on extensive collection of real-world received signal strength (RSS) fingerprints, making deployment costly and time-consuming. We present a ray-tracing-based framework that reduces this reliance by generating synthetic RSS fingerprints from a building model. 
We first calibrate the building model using a small amount of real RSS fingerprints through Bayesian optimization, followed by per-access-point calibration to account for residual errors in simulated RSS values. The calibrated model is then used to generate a large augmented dataset of synthetic RSS fingerprints at arbitrary locations. To effectively exploit these data for localization, we introduce novel binary and multivalued representations of RSS values and a ResNet-based localization architecture that supports cross-band fusion of 2.4 and 5 GHz measurements. We evaluate our localization method on a real campus building against a diverse set of four baselines. When trained exclusively on synthetic data, the proposed method with multivalued representation and upstream cross-band fusion achieves a mean localization error of 3.05m on a real-data test set, outperforming the best baseline by 33.6\%. The results demonstrate that calibrated ray-tracing-based simulation can substantially reduce the need for real RSS fingerprints while enabling accurate deep-learning-based indoor localization.
\end{abstract}

\maketitle

\section{Introduction}
\label{sec:introduction}

The persisting appeal of using signal strength measurements from Wi-Fi access points (APs) for the purposes of indoor localization is due to three factors: (a) today, Wi-Fi APs are ubiquitous in indoor environments, and often deployed in large numbers, (b) Wi-Fi-enabled devices, like smartphones, are already equipped with the hardware to collect the necessary Received Signal Strength (RSS) measurements, hence no additional hardware cost is involved, making RSS readings a ``least common denominator'' of capability present over a large swath of devices, and (c) RSS-based localization algorithms and products \cite{bahl_radar_2000,hart_how_2025,_cisco_} have been demonstrated, reinforcing the confidence that, as long as an application or the user find acceptable a localization error of a few meters, RSS-based localization can be adequate. Based on those observations, it would appear that RSS-based localization is a ``solved'' problem. 

While RSS-based localization still evolves -- indeed we propose new localization methods in the current paper -- its main shortcoming is a, largely, engineering problem. The cost of RSS-based localization is the (one time) cost of the of pre-deploying APs -- usually ignored because the APs are already present for their data communication services -- and the cost for a number of measurements (``samples'' or ``fingerprints'') necessary to collect using some form of site survey. The engineering challenge is that, when the wireless propagation environment changes, perviously collected fingerprints are no longer relevant and new ones may have to be collected, thus forcing a repetition of the site survey, introducing a recurring cost. Note that one could, at least in principle, deal with the changing environment by inflating the density of deployed APs, making them act as ``proximity sensors'' so no matter what happens to the environment, a device is always close to an AP, whose coordinates then become an approximate location of the device. This approach is generally considered  unacceptable due to the cost of equipment and wiring. The sensible alternative is re-running site surveys, and therefore the problem at hand is that of reducing the effort of the site survey or fingerprinting. 

In this paper we address the question of expanding a small set of ground truth fingerprints by means of a data augmentation process that allows the generation of synthetic fingerprints. It does not matter whether we are addressing the initial site survey, or a subsequent one when the environment changes -- the process is the same. Our work hinges, in addition to factors (a)-(c) mentioned in the beginning of this introduction, on a fourth factor: (d) an increasing number of indoor environments are described by 3-dimensional building models, readily available for a variety of tasks, e.g.,  for the definition and operation of ``digital twin'' representations. The 3-dimensional building models, with the additional information of the AP locations,  can be used to inform radio frequency (RF) propagation modelling software, and ray tracing tools in particular, to derive estimates of signal strength measurements across space. 

We describe how synthetic ray-traced set of RSS values are aligned to a small set of ground truth RSS fingerprint measurements via an automated calibration process which operates at two levels. The first calibration adjusts the parameters of the building materials used in the propagation model of the ray tracer such that the produced values align with the measured ground truth. This is particularly useful in legacy buildings, such as the example one used in the evaluation section of this article, where there is a degree of uncertainty on the exact composition of the various building materials used in their construction and subsequent lifetime of renovations. The second calibration adjusts the per-access-point measurements from the ground truth data set to account for residual errors in the RSS values synthesized by the ray tracer. Thus, our calibration handles both building model uncertainty and the approximate nature of ray tracing results. 

Additionally, given the pervasive use of two ISM bands (the 2.4GHz and the 5GHz ISM bands) for Wi-Fi communication, and the simultaneous operation of today's APs on both bands, we introduce \emph{fusion} of fingerprints (ground truth and synthesized) from the two bands and study its impact on localization accuracy. The rationale behind this fusion is that, for the same environment, different radio frequencies encounter different propagation characteristics as the interaction with the materials of the buildings is frequency-specific. Hence, at each location, two, diverse, RSS measurements, one for each frequency, represent two different measures of the impact of location on the received signal. The introduced diversity can be generalized to more frequency bands as they become more widely used; e.g. the 6GHz band introduced since IEEE 802.11ax (``Wi-Fi 6E''). The necessary pipeline for synthesizing fingerprints remains the same, independently, and separately, synthesizing RSS values for each of the used bands. 

In addition to determining the impact of the proposed augmentation method, we also propose and evaluate a new localization algorithm based on the use of 2-dimensional heatmap ``images'' -- one for each AP -- that corresponds to the RSS values measured (and synthesized) across space. The images are treated as input channels fed into a deep convolutional neural network (CNN) architecture which predicts the location of the Wi-Fi-enabled device in the 2-dimensional plane. Our evaluation section provides results for the new localization scheme, as well as for four baseline methods, two learning-based methods including a state-of-the-art, and two non-learning based ones, one geometrically-inspired using intersection of convex hulls and one using density-based clustering. For the non-learning ones, the noisy and imperfect nature of RSS measurements is handled via binning of the RSS values, independently for each AP. 

Our contributions are summarized below:
\begin{itemize}
\item We develop a method for augmenting RSS measurements using ray tracing, a building information model~(BIM)~\cite{_bim_}, and the locations of Wi-Fi APs. We further propose an automated two-step calibration method to align the augmented measurements with ground-truth data.
\item We introduce a novel transformation that converts both simulated and real RSS measurements into per-AP heatmap images. Using this representation and the augmented data, we train a modified ResNet-18 model to estimate device location. We also propose a data fusion approach that combines RSS measurements from multiple frequency bands, when available, to improve localization accuracy.
\item We conduct an extensive evaluation of the proposed localization method on a floor of a campus building, comparing its performance against four baselines, including the state-of-the-art.
\end{itemize}
The evaluation results demonstrate the benefits of the proposed data augmentation scheme, and the newly proposed localization approach improves accuracy against the state-of-the-art by 1.54 meters for the indoor environment used in our evaluation section. 

The remainder of the paper is structured as follows: Section \ref{sec:related_works} covers related work, emphasizing the existing approaches to data augmentation applicable to RSS data. Section \ref{sec:methodology} covers the data augmentation methodology, including the calibration steps, and also introduces a new localization method. Section \ref{sec:setup} details the experimental setup, and describes the localization baselines against which the new localization scheme is compared. Section \ref{sec:results} presents the results confirming the benefits of data augmentation, the impact of fusion of RSS from different frequencies, and the overall improvement possible against baseline schemes. We conclude in Section \ref{sec:conclusions} with a summary of the main conclusions and the steps for future work.

\section{Related Work}
\label{sec:related_works}
Wi-Fi fingerprinting is a localization technique that estimates the location of a Wi-Fi-enabled device 
by matching its observed signal characteristics to a previously constructed database (\emph{RSS or radio map}) containing measurements collected at known locations.
Since the early 2000s, a significant amount of research has been conducted in this area~\cite{mostafa_survey_2025}, 
leading to improvements in localization accuracy, robustness, training efficiency, and deployability.
The very first work was RADAR~\cite{bahl_radar_2000}, a framework that predicts the location of a mobile device in an office environment using RSS measurements from three Wi-Fi APs.
RADAR derives a location estimate by finding the closest RSS measurement in the fingerprint database.
Subsequently, MoLoc \cite{sun_moloc_2013} sought to use motion data in conjunction with Wi-Fi fingerprinting to identify and discount noisy measurements, thereby improving localization performance.
Building on similar ideas, Han et al.~\cite{han_building_2014} developed a framework using weighted k-nearest neighbours and multiple filters to build a practical indoor navigation system relying solely on Wi-Fi.
As an example of current, state-of-the-art commercial offerings, we note a system called \emph{Hyper} that claims 1-meter indoor localization accuracy by combining Wi-Fi fingerprints with accelerometers, gyroscopes, and camera systems~\cite{_hyper_}.
All these approaches require an initial site survey that is both lengthy and physically demanding, as the entire space needs to be traversed and RSS measurements be taken at many locations to build an accurate RSS map of the environment. 
A crucial factor for the success of Wi-Fi fingerprinting schemes is therefore the reduction of the effort required for site surveys.

\subsection{Neural Network-Based Localization with Fingerprinting}

RSS fingerprint data, labelled by the corresponding known locations, can be used in supervised training of models that, given an RSS fingerprint as input can produce location information. On this principle, a number of attempts have been made to train a neural network on Wi-Fi RSS fingerprints.
For instance, Ibrahim et al. \cite{ibrahim_cnn_2018} proposed feeding the RSS data as a time-series input to a CNN. This model achieved an accuracy of 100\% in building and floor prediction, and a localization error of 2.77 meters combining  10 successive most recent RSS records as input.
However, its long inference time of up to 60 seconds limits real-world deployments, and its mean localization error increases to 10.25 meters if it receives a single RSS record as input.
Song et al. \cite{song_novel_2019} proposed another CNN-based approach called CNNLoc where a stacked autoencoder is used to extract certain features from raw RSS fingerprints; these features are then fed to the CNN to predict a user's location.
It achieved an accuracy of 100\% in building classification and 95\% in floor classification, and the mean localization error was between 7.6 and 11.78 meters across 3 different datasets.
The large localization error of over 7 meters limits the prospect for real-world deployments.
Laoudias et al. \cite{laoudias_localization_2009} used a Radial Basis function (RBF) network to estimate a user's location from RSS measurements and achieved a mean location error of 3.4 meters.
In addition to the localization error magnitude, another key problem that hampers the development of the above mentioned localization techniques is the amount of high quality fingerprint samples required to train these neural networks. 
This highlights the need to augment the training dataset without doing an extensive site survey. 

\subsection{Wi-Fi RSS Data Augmentation}
Data augmentation is the task of expanding the set of ground truth measurements with synthesized values that are ``as good as'' the real measurements. In this section, we narrow our attention to augmentation applicable to Wi-Fi fingerprinting (see a recent survey~\cite{feng_survey_2025}). 
We review and discuss different categories of augmentation methods below.

\subsubsection{Traditional Augmentation}
Traditional augmentation methods operate directly on collected RSS fingerprint data and 
generally have much simpler modelling or learning procedures.
These approaches typically rely on statistical manipulation, interpolation, or perturbation 
of existing measurements to generate additional samples for learning.
Due to their simplicity and low computational requirements, traditional methods 
are still wildly used in many RSS fingerprinting methods.

One commonly used technique is resampling and permutation, 
as demonstrated by Sinha and Hwang~\cite{sinha_improved_2020}.
The authors propose a resampling strategy in which RSS values from different measurements at the same location are randomly combined to create new synthetic training samples alongside the original data.
This resampling method preserves the statistical distribution of the original dataset while significantly increasing the number of fingerprint samples available for training.
However, it only generates new samples for locations they have measurements from, and cannot be used to estimate non-measured locations.

Another approach is perturbation, where small random variations are added to the RSS values in order to simulate real-life variations driven by environmental fluctuations, device variability, and measurement noise.
An example of this can be seen in the work of Sinha et al.~\cite{sinha_data_2019}, where two perturbation-based augmentation schemes are proposed: 
one adjusts a single RSS value by a fixed constant (in this case $-5$) across $N$ copies, where $N$ is the number of visible access points at the measurement location, while the other produces uniformly random values between the mean and the original RSS value for each measurement location.
However, perturbation-based augmentation suffers the same issues as resampling, where it cannot generate samples for locations it has not measured from. 

In interpolation-based approaches, fingerprints at locations with no ground truth measurements are generated by estimating the RSS values from nearby known measurement locations.
These techniques differ from the previous two in that new ``virtual'' measurement \emph{locations} are introduced at which the localization model can make a more precise estimate.
There are several interpolation techniques that can be used; one of them used a quadratic polynomial fitting to model the relationship between RSS values and distance \cite{yang_semisimulated_2020}, another used a cubic spline and both 2.4 and 5 GHz frequency bands to improve localization \cite{_indoor_}, a more sophisticated employed by Zhao et al. \cite{zhao_applying_2016}, is \emph{Kriging interpolation}.
In this paper, the authors increase the number of locations by modelling spatial drift and RSS variance based on the degree of spatial dependence of the dataset and how correlations decay with distance.
However, interpolation-based augmentation methods of this kind assume smooth spatial variation in RSS values, which may not be the case in complex indoor environments.

\subsubsection{Propagation Model-Based Augmentation}
Propagation-based modelling techniques attempt to address the limitations of traditional methods by incorporating a physics-based, analytic approach to RF propagation.
Instead of using interpolation techniques to increase the density of measurement locations, these approaches use analytical propagation models to estimate RSS values at unmeasured locations.  

One commonly used model for this approach is the \emph{log-distance path loss (LDPL)} model, which describes how signal strength decreases with distance from the transmitter.
This model was used by He et al. \cite{he_novel_2016} to generate RSS maps for indoor localization without performing extensive fingerprinting measurements.
The authors employ the log-distance path loss model 
with an additional Gaussian random variable term 
to model signal attenuation caused by shadow fading.
The LDPL model is given below
\begin{equation}
    \label{eq:loss-distance-path-loss-model}
    P(d) = P_0 - 10\alpha\cdot \log_{10}\left(\frac{d}{d_0}\right) + X_r,
\end{equation}
where $d$ is the distance between the AP and the target location,
$P_0$ is the reference power of the Wi-Fi signal at distance $d_0$ from the AP, 
$\alpha$ is the path loss exponent due to indoor space characteristics, and
$X_r$ is a Gaussian random variable to mimic shadow fading.
By calibrating the path-loss parameter $\alpha$ using ground truth RSS measurements, 
the model can estimate RSS values throughout the indoor environment.

To further improve localization in more complex environments with multiple walls and floors, one can use a multi-wall propagation model, as employed in \cite{lemic_enriched_2016}, where signal attenuation is modelled as a function of both distance and environmental obstructions such as walls and floors.
The model used in that work is
\begin{equation}
    \label{eq:multiwall-propagation-model}
    L(d) = L_0 + 10 \alpha_{\text{MW}} \log_{10}(d) + \sum_w \beta_w W_w + \sum_f \beta_f F_f ,
\end{equation}
where $d$ is the distance between the AP and the target location,
$L_0$ is the reference loss at $d_0$,
$\alpha_{\text{MW}}$ is the path loss exponent for the multi-wall model,
\(\beta_w \) and \(\beta_f\) are the attenuation per wall and floor respectively, and
\(W_w \) and \(F_f \) are number of crossed walls and floors respectively.
By accounting for environmental obstructions like walls and floors, multi-wall models can generate more realistic RSS fingerprints compared to simpler distance-based models.
 
Guan et al. \cite{guan_robust_2015} proposed REAL for robust, cost-effective, and scalable localization in large indoor areas.
REAL employs a probabilistic propagation model that takes into account both wall attenuation and distance.
The model used in that work is
\begin{equation}
    \label{eq:REAL_model}
    P(d) = P_0 + \alpha_{\text{REAL}} \log_{10}(d) + \sigma_{\text{REAL}}N_{ob} + X_\epsilon,
\end{equation}
where $P_0$ is the Wi-Fi RSS value at 1-meter distance from the AP,
$\alpha_{\text{REAL}}$ is the attenuation factor due to distance,
$\sigma_{\text{REAL}}$ is the attenuation factor due to walls,
$N_{\text{ob}}$ is the number of walls on the path of the signal, and
$X_{\epsilon}$ is the Gaussian modelling error.
The calibration of REAL depends on the size of the available real training data: for small training sets, it assumes a homogeneous set of APs (i.e., all APs share the same $P_0$ and $\epsilon$ parameters), while for larger training sets, it models each AP separately, capturing the differences that better represent each AP's actual environment.
 
In general, propagation-based modelling methods can quickly generate Wi-Fi fingerprints for a large area, eliminating the need for the expensive and time-consuming step of manually creating an RSS map. Their main shortcoming is that they are just analytical approximations of a very complex phenomenon, that of wireless propagation. They therefore exhibit poor generalizability properties when tried across different environments. While these models can be, and usually are, informed by data specific to an environment, they are only informed by means of broad statistics, like averages. Subsequent techniques have looked more closely into extracting more information from the measurements collected in an  environment, to build regression models (\ref{regression}), and for training generative models (\ref{gan}).

\subsubsection{Regression-Based Augmentation} \label{regression}%
Supervised learning approaches form another class of techniques that is rapidly growing in interest.
Rather than relying exclusively on analytical models, these approaches learn the statistical relationships between spatial locations and RSS measurements directly from collected data.
Once trained, the model can estimate RSS values at unmeasured locations, generating new fingerprints to support more accurate localization.
 
One of the most widely used techniques in this category is \emph{Gaussian Process Regression (GPR)}.
Such models estimate the RSS value as a random variable drawn from a Gaussian process defined by a covariance function that captures spatial correlations.
This technique was used by Sun et al.~\cite{sun_augmentation_2018} to generate an RSS map from a limited set of labelled fingerprints.
To improve localization accuracy, compound kernels, such as Matern and Rational Quadratic, were combined to capture both smooth trends and local variations.
Support Vector Regression (SVR) has also been used to estimate RSS values at unmeasured locations by Mendoza et al. \cite{mendoza-silva_environmentaware_2022}.
The authors used a linear kernel function to learn the spatial and environmental features of the APs' deployment.

In recent work, a neural network has been used to augment the fingerprint data~\cite{lan_fingerprint_2022}. The authors convert the sparse Wi-Fi fingerprint data into a low-resolution image before being processed into a super-resolution CNN model to produce a high-resolution image. These high-resolution images are then down-sampled and converted back into an augmented Wi-Fi fingerprint database.  

\subsubsection{Augmentation via Generative Models} \label{gan}
Generative Adversarial Networks (GANs) are a class of deep learning models that have been used to augment training data in order to improve model performance~\cite{bowles_gan_}.
They are comprised of two models trained simultaneously: a generator that produces synthetic data to mimic the training data, and a discriminator that attempts to distinguish the training data from the generated (synthetic) data \cite{goodfellow_generative_2014}.
The generator is trained to maximize the discriminator's probability of classifying generated samples as real (or equivalently, to minimize its adversarial loss), 
causing the generated data distribution to approach the distribution of the training data.
GANs have been used by Junoh et al. \cite{junoh_enhancing_2024} to generate synthetic data for unmeasured locations based on sparse crowdsourced data.
In addition, conditional GANs (cGANs), which are a variant of GANs,
can use annotated information to provide additional context to the generated data.
Such additional information can include building or floor IDs, used by both the generator (to provide region-specific synthetic data) and the discriminator (to evaluate the generated data with the added context).
Because cGANs allow location-aware data generation, they are particularly useful in large, diverse, and complex environments where the generator can learn the nuances of different sub-environments.
This can be seen in \cite{quezada-gaibor_surimi_2022}, where a cGAN was used with the additional label of the building and floor ID providing additional context to both the generator and the discriminator.

Finally, a composite approach was proposed in \cite{yean_extendgan_2023} combining Dirichlet distribution-based upsampling and Wasserstein GAN with Gradient Penalty (WGAN-GP) in a framework called extendGAN+.
The authors first train a base WGAN-GP model using the existing training data, then transfer and fine-tune the model using transfer learning to generate synthetic RSS data for unmeasured locations.
To further improve the quality of the generated data, filtering was used to eliminate the outliers based on dissimilarity thresholds, then the final RSS values where transformed into a 2D feature vector, which was subsequently processed by a deep convolutional neural network to predict the user's location.

\subsection{Ray-Tracing for RF Modelling}

Rather than investing effort on extracting ever more elaborate models from a few measurements, an alternative is to invest effort in describing the environment and linking the RSS information to the properties of the materials in the environment. 
The wave nature of RF and visible light propagation, leads to the use of similar tools in both domains.  \emph{Ray tracing}---a technique originally developed for realistic illumination rendering in modern video games and computer-generated films---has been used to simulate realistic RF propagation within an environment.  Therefore ray tracing can be employed to reduce or eliminate the need for site surveys. 
An early example of RF ray tracing comes from work published in 2015~\cite{yun_ray_2015}, with more recent work including modelling surface scattering and reconfigurable intelligent surfaces (RISs) \cite{choi_withray_2023}.
Beyond signal strength estimation alone, the use of ray tracing to model RF propagation can additionally provide features like angle of arrival, angle of departure, phase information, and time of flight, all of which carry potentially useful information for localization tasks. 

A caveat is that ray tracing relies on how accurate the propagation environment is described and modelled. In most realistic settings, there is limited fidelity in how the propagation environment (walls, windows, objects, etc.) is modelled, meaning the ray tracing computation produces approximate results. Moreover, ray tracing is computationally expensive but modern Graphics Processing Units (GPUs) have become capable of performing even real-time ray tracing computations of adequate fidelity. 
The most recent and advanced software for ray tracing-based RF propagation that we have identified is \emph{Sionna-RT}~\cite{hoydis_sionna_2023}.
More ray tracing RF simulators exist (Opal \cite{egea-lopez_opal_2021}, NYURay \cite{kanhere_calibration_2025}, and PyLayers \cite{amiot_pylayers_2013} inter alia) but they all have limitations that Sionna RT does not have.
For example, modelling environments in PyLayers is less expressive and more cumbersome than that in Sionna RT, while NYURay has a comparable feature set of model expressiveness but does not produce a radio map needed for our work.
Opal, which is the closest to Sionna RT in terms of feature set also does not have an easy method to produce a radio map.
In this work, we have selected Sionna RT as it provides a straightforward process of converting BIM files into a digital model for RF propagation simulation and is, at the time of writing, actively supported and evolved by its developer community. 

Employing ray tracing as a data augmentation strategy of RSS measurements requires a means to align ray tracing output to ground truth RSS measurements to counter the approximation introduced by limiting the execution time of ray tracing as well as the effects of the imperfect description of the environment and the properties of RF transmitter and receivers. We address those issues in the current paper. 

\section{Methodology}
\label{sec:methodology}
\begin{figure}[ht]
  \centering
  \includeinkscape[width=0.65\linewidth]{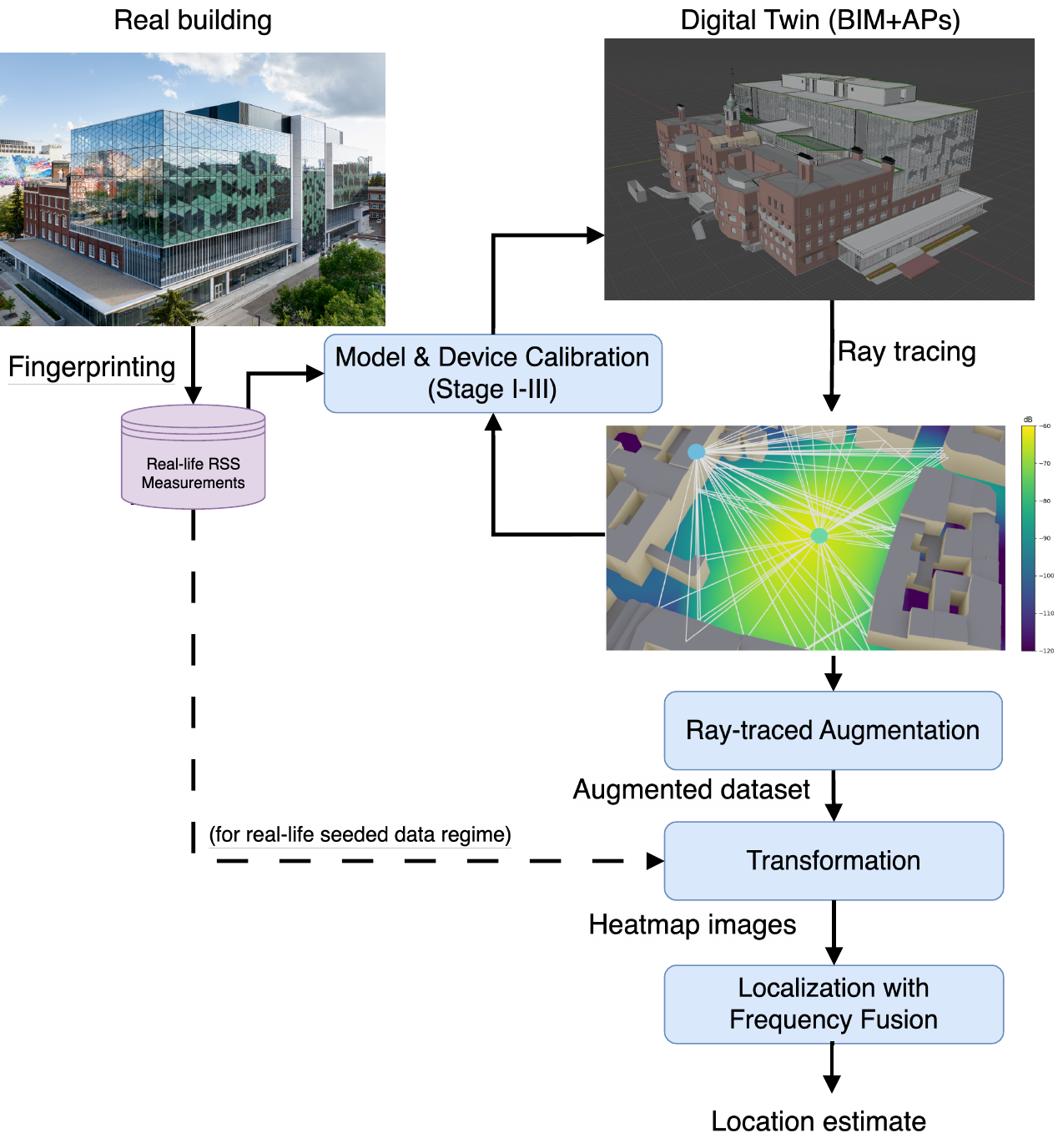}
  \caption{Illustration of the proposed methodology for indoor localization}
  \label{fig:overall-pipeline}
\end{figure}

Our localization framework consists of two main components: signal-strength data augmentation and a learning-based localization method that uses the augmented data to estimate the location of an individual carrying a Wi-Fi-enabled device.
\Cref{fig:overall-pipeline} shows the overall pipeline, which begins by modelling the building environment to generate a BIM file. The BIM is then used as a digital twin to calibrate Sionna RT for ray tracing. The calibrated simulator is subsequently used to augment the RSS data, to be used in the subsequent localization steps.

\subsection{Ray-Traced RSS Augmentation}
To train an expressive neural network for localization, 
we first need to augment our limited, real-life dataset. As an example, in our evaluation section we operate using only 34 locations of ground truth data and augment them using Sionna RT and the process described to produce 2,155 location measurements. 
The process involves modelling the indoor environment, calibrating both for building model and RF transmitter/receiver uncertainty, and finally generating synthetic RSS data for training.

\subsubsection{Modelling the Built Environment}
To model our environment we used   \emph{Blender} \cite{foundation_blender_} to import the Building Information Model (BIM)~\cite{_bim_} file that represents the real environment.
The BIM file is assumed to accurately represent the building's floor plan and geometry, including the locations of doors, windows, and interior and exterior walls.
Once the building has been loaded, material properties are assigned to the objects in the scene based on observations made during the site survey.
Sionna RT provides a collection of common material types with known RF properties, which we use to assign material parameters in our model.
When a specific material is not included in this collection, we select the closest available alternative based on its composition or function.
For example, because carpet is not among the predefined materials, we use floorboard as the closest match;
while Sionna does permit custom materials to be defined, doing so requires additional testing and equipment, which lies outside the scope of this work.
To mitigate the impact of material mismatches,  we apply a calibration process described in \Cref{sss:sionna_calibration}.
In cases where the BIM file does not specify the placement of the Wi-Fi APs, 
we manually add them within Blender based on observations and floor plans.
Sionna RT uses these AP objects as transmitters to generate the radio map to augment our data. 
Additionally, each AP object is labelled to allow easy lookup of the networking properties of the corresponding real AP.

\begin{figure}[h!]
    \centering
    \includegraphics[width=0.6\linewidth]{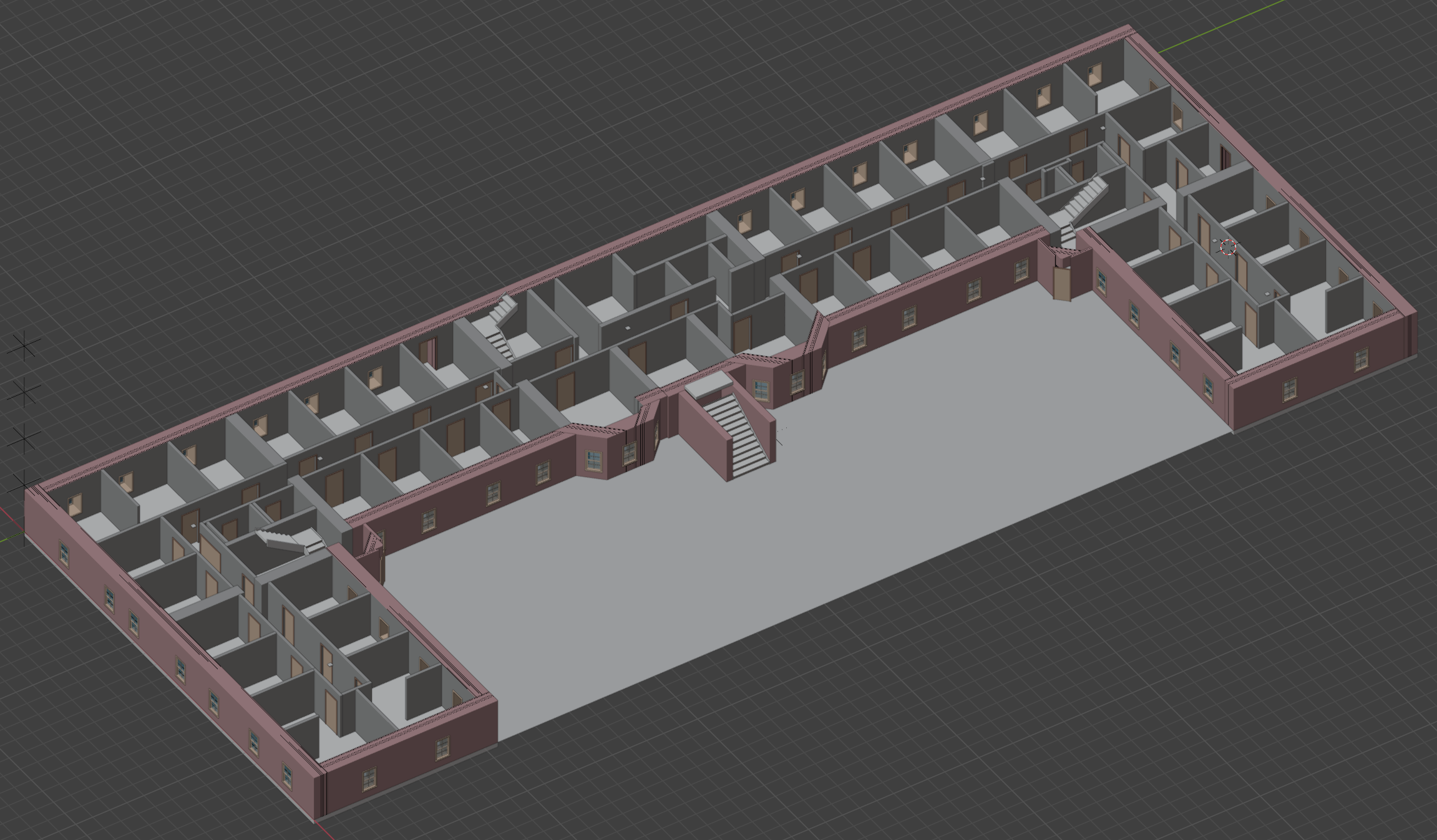}
    \caption{Orthographic view of Level-1 of our campus building}
    \label{fig:ATH-level-1-ortho-blender}
\end{figure}

\Cref{fig:ATH-level-1-ortho-blender} shows an orthographic view of the first floor of the university campus building that we use in this study, with the ceiling and floors above hidden so that the processed model can be seen with all of its assigned materials in place.
Different shaders and colors are used to visualize which materials have been assigned to which objects.
Before exporting, all floors and physical objects are made visible again, so that they are seen by Sionna RT.
The final processed model is then exported in the Mitsuba file format for import into Sionna RT.

\subsubsection{Calibration}
\label{sss:sionna_calibration}
Calibrating our digital model to allow simulation of RF propagation involves three stages, all of which are important as they affect the simulated RSS values on which fingerprint-based localization ultimately relies. We outline these stages below.

\textit{Stage I:} Material types and thicknesses are not often captured in BIM models accurately, especially for legacy buildings, owning to renovations, imprecise modelling, poor documentation, and other imperfections. 
To ensure that Sionna RT can accurately model RF propagation throughout the environment, we first calibrate the material parameters (i.e. thicknesses) in the BIM model. 
Specifically, we use Bayesian optimization (BO)~\cite{snoek_practical_2012} to identify the material parameters that minimize the average error between simulated and measured RSS values at specific calibration locations.
This objective function, denoted as $f(x)$, can be written as:
\begin{equation}\label{eq:mse}
f(\mathbf{x}) = \frac{1}{|\mathcal{L}|} \sum_{l\in \mathcal{L}} \left( \text{RSS}_{\text{pred}}(l, \mathbf{x}) - \text{RSS}_{\text{true}}(l) \right)^2
\end{equation}
where $\mathcal{L}$ is the set of calibration locations, $\text{RSS}_{\text{true}}(l)$ is the average of all RSS values measured at location $l$ during the data collection period, and $\text{RSS}_{\text{sim}}(l, \mathbf{x})$ is the simulated RSS value at that location, which depends on the material thickness vector $\mathbf{x}$. 
Evaluating $f(\mathbf{x})$ at each $\mathbf{x}$ in order to obtain $\text{RSS}_{\text{sim}}(l, \mathbf{x})$ is computationally expensive to perform with ray-tracing. For example, for the campus building used in our evaluation section, and 10 APs, it takes an average of 30 seconds to complete the RF ray-tracing for all APs and compute $\text{RSS}_{\text{sim}}$ for a specific material parameter vector.  
BO efficiently optimizes this objective while requiring relatively few evaluations compared to exhaustive search or greedy local search.

The average discrepancy is indeed a \emph{black-box function} of the material parameters, denoted as $f: \mathbb{C} \rightarrow \mathbb{R}$ where $\mathbb{C}\subset \mathbb{R}^d$ 
is a compact set and $d$ is the number of material parameters that must be tuned.
To reduce the number of expensive function evaluations, BO approximates the black-box function using a probabilistic \emph{surrogate model} that is iteratively refined as new observations (i.e. simulation results) become available. 
The next evaluation point is then selected by maximizing an \emph{acquisition function} that
balances exploration (sampling regions with high uncertainty, or high potential for improving the objective, etc.) and exploitation (sampling regions predicted to have low objective values). 
In our implementation, we use a Gaussian Process (GP) as the surrogate model, the Max-value Entropy Search (MES)~\cite{wang_maxvalue_2018} as the acquisition function, and a randomly selected point as the first evaluation point. 
MES selects the next evaluation point, $\mathrm{x}_{n+1}$, by maximizing the expected information gain about the global optimum of the black-box (objective) function: 
\begin{equation*}
  \label{eq:MES}
  \mathrm{x}_{n+1} = \mathop{\mathrm{argmin}}_{\mathrm{x}\in\mathbb{C}} \mathrm{I}((\mathrm{x}, \mathrm{z});\mathrm{z}^* |\mathcal{D}_n)=
  \mathop{\mathrm{argmin}}_{\mathrm{x}\in\mathbb{C}} \mathrm{H}(P(\mathrm{z}|\mathcal{D}_n, \mathrm{x})) - \mathbb{E}[\mathrm{H}(P(\mathrm{z}|\mathcal{D}_n, \mathrm{x}, \mathrm{z}^*))]
\end{equation*}
with $\mathrm{I}(.;.)$ being the mutual information, $\mathrm{H}(.)$ being the entropy, $\mathrm{z}^*=f(\mathrm{x}^*)$ being the minimum value of $f$ on its domain, 
and $\mathcal{D}_n=\{(\mathrm{x}_\tau, \mathrm{z}_\tau)\}^n_{\tau=1}$ being our observations for the first $n$ evaluation points. 
We empirically found that MES consistently outperforms other commonly used BO acquisition functions, such as expected improvement, probability of improvement, and lower confidence bound, for this calibration problem.

\textit{Stage II:} We generate the radio map once the model calibration is complete.
First, the calibrated material thicknesses are applied to objects that make up the digital model.
Transmitters are then placed at the AP locations specified by the model; 
their parameters are configured to match those of their real-world counterparts, including transmit power, antenna pattern, orientation, bandwidth, and operating frequency.
After all the parameters are defined, we generate the radio maps for each AP within the environment.
Specifically, we run the radio map solver in Sionna RT multiple times to produce a set of radio maps at different heights.
Generating radio maps at multiple heights not only increases the number of samples available, but also captures variability that arises from the fact that real measurements may not have been taken at the exact height assumed in the simulator.
These radio maps are subdivided into equal-sized rectangular cells called \emph{measurement cells}, which aggregate all the path losses measured within the cell into a single value.
We then transform the path loss map into an RSS map by multiplying the path loss by the transmit power, yielding RSS values in milliwatts (mW): 
\begin{equation}
  \label{eq:path_loss_to_RSS}
  \text{RSS}_{i, tx} = PL_{i, tx}\times P_{tx}  
\end{equation}
\noindent
where $\text{RSS}_{i,tx}$ is the received power in mW at location $i$ for transmitter $tx$, $PL_{i, tx}$ is the path loss to measurement cell $i$ from that transmitter, and $P_{tx}$ is the transmit power in unit of milliwatts for that transmitter. 

Since most devices report RSS in decibel-milliwatts (dBm), 
RSS values are converted from mW to dBm using the following equation:
\begin{equation}
  \label{eq:RSS_mw_to_dbm}
  \text{RSS}_{\text{dbm}} = 10 \times \log_{10}(\text{RSS}_{\text{mW}})
\end{equation}
\noindent
At this point, the converted radio maps could, in principle, be used to estimate the RSS value at an arbitrary location. 

\textit{Stage III:} The final stage is device calibration which corrects for residual errors arising from phenomena that are not adequately modelled by Sionna RT.
These include RF losses within the receiver and transmitter caused by manufacturing inconsistencies, humidity, antenna gain variations, and similar factors.
For device calibration, we compute the average difference between the simulated and real RSS values at the corresponding locations for each Wi-Fi AP.
This difference yields a calibration offset, $\Delta_{\text{RSS}}(\text{AP})$, which translates real measurements into simulated measurements that Sionna RT would have produced if the unmodelled losses had been accounted for.
The offset does not fundamentally alter the signal loss curve used to determine the signal strength at a given location; 
rather, it shifts the curve relating real to simulated measurements by an amount equal to the sample-average difference.
This offset can be calculated as follows:
\begin{equation}
\Delta_{\text{RSS}}(\text{AP}) = \frac{1}{|\mathcal{L}_{\text{AP}}|} \sum_{l \in \mathcal{L}_{\text{AP}}} \frac{1}{|\mathcal{M}_{\text{AP},l}|} \sum_{m \in \mathcal{M}_{\text{AP},l}} \left( m - \text{RSS}_{\text{real}}(\text{AP}, l) \right)
\label{eq:ap_offset}
\end{equation}
where $\mathcal{L}_{\text{AP}}$ is the set of valid measurement locations for which the ground truth RSS values are finite for $AP$,
$\mathcal{M}_{\text{AP},l}$ is the set of up to $K$ valid simulated measurements drawn from $K$ RSS maps measured from $K$ different heights for the transmitter $\text{AP}$ at location $l$, and $\text{RSS}_{\text{real}}(\text{AP}, l)$ is the ground truth RSS value for $\text{AP}$ at location $l$.
The computed offset is then used to translate between the real RSS values and the simulated RSS values from Sionna RT.

Rather than applying the computed offset directly to the generated RSS maps, we leave the maps unmodified and instead use the offset to translate the real RSS values into their expected simulated counterparts.
This choice allows us to support multiple devices, each with its own calibration information, without having to generate a separate, device-specific RSS map of the same environment.
Since the calibration consists of a simple per-AP constant offset between the real and simulated values, conversion in either direction is given directly by \Cref{eq:get_rss_sim_from_rss_real}:
\begin{align}
\label{eq:get_rss_sim_from_rss_real}     
    \text{RSS}_{\text{sim}}(AP) &= \text{RSS}_{\text{real}}(AP) + \Delta_{\text{RSS}}(\text{AP})
\end{align}
\noindent

\subsubsection{Data Augmentation}
\label{sss:data_auggmentation}
With the calibrated RSS maps in hand, we now explain how they are used to augment our real RSS dataset.
Recall that $K$ RSS maps are generated at $K$ different heights, and within an RSS map, each measurement cell contains a vector of RSS estimates from all APs in range.
We use these RSS maps for data augmentation.
For each measurement cell whose entire footprint lies within a room or hallway,
we use the estimated RSS values from the $K$ RSS maps to augment the set of available samples. Each sample consists of an RSS value and a label representing the $x$ and $y$ coordinates of the centre of the corresponding measurement cell.
Thus, the augmentation process yields the following dataset for each measurement cell $\Big\{\big(\text{RSS}_{\text{sim}}^{(1)}(AP_i),(x,y)\big),\dots,\big(\text{RSS}_{\text{sim}}^{(K)}(AP_i),(x,y)\big)\Big\}_{\forall AP_i}$ with $(x,y)$ being the centre of that cell,
and the final synthetic dataset is the union of these sets over all valid measurement cells.
This data augmentation process is separately performed on the 2.4 and 5 GHz bands. 

\subsubsection{Transformation}
\label{sss:transformations}
Localization techniques require different input formats. Some methods directly use the vector of RSS values, whereas others operate on representations derived from these values. To accommodate a range of localization techniques, we define three transformations of the samples in the augmented RSS dataset described in \Cref{sss:data_auggmentation} to produce three different representations: (1) an identity transformation that directly returns the vector of RSS values at a given location; (2) a transformation that maps the RSS vector to a binary matrix, analogous to a black-and-white image, per AP; and (3) a transformation that maps the RSS vector to a multivalued matrix, analogous to a grayscale image, per AP. We describe each data representation and the corresponding transformation below.

\paragraph{RSS Vector}
\label{sss:data_gen_method_rss_fingerprinting}
The first input representation is an RSS vector corresponding to a measurement cell in the RSS map. Since the augmented dataset contains RSS vectors already, the required transformation is simply the identity map. This representation is used primarily for comparisons with related work in which localization techniques rely solely on the measured RSS values to estimate the user's location.

\begin{figure}[ht!]
    \centering
    \hfill
    \begin{subfigure}{0.46\textwidth}
        \centering
        \includeinkscape[width=\textwidth]{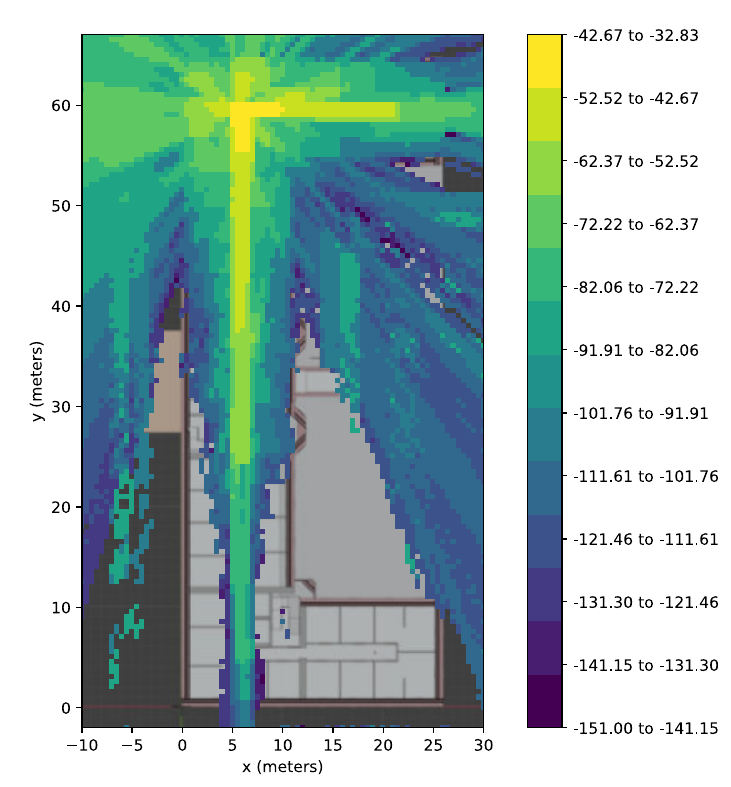}
        \caption{Binned RSS values for AP\_002 (the bin size is 9.85 dBm)}
        \label{fig:rss_bins_ap_002}
    \end{subfigure}
    \hfill
    \begin{subfigure}{0.52\textwidth}
        \centering
        \includeinkscape[width=\textwidth]{svg-inkscape/AP_002_selected_bin_svg-tex.pdf_tex}
        \caption{Selected measurements cells for AP\_002 given the reference RSS value of -75.0 dBm}
        \label{fig:rss-select-bin-ap_002}
    \end{subfigure}
    \caption{Binning then selecting a bin based on a reference value for AP\_002}
    \hfill
\end{figure}

\begin{figure}[h!]
    \centering
    \includeinkscape[width=0.4\textwidth]{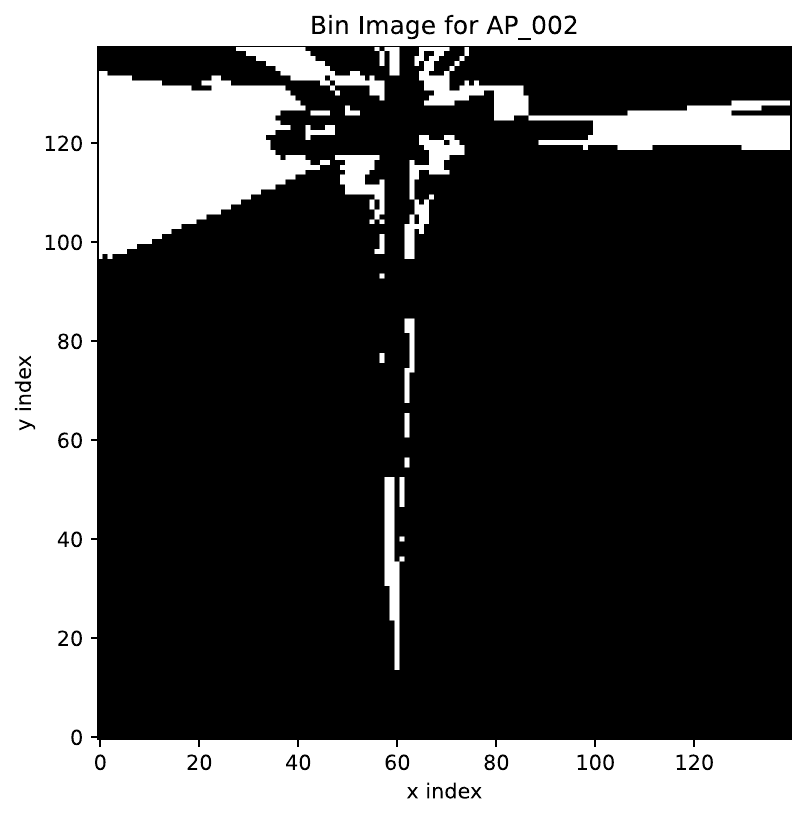}
    \caption{An example of a binary matrix/image constructed from RSS values for AP\_002}
    \label{fig:ap_002_bin_image}
\end{figure}

\paragraph{Binary Matrix}
\label{sss:data_gen_method_rss_bin_maps}
The second input representation is a binary matrix for each AP, where each element corresponds to a measurement cell. An element is set to 1 when the average RSS value of the corresponding measurement cell is sufficiently close to the measured RSS value, and to 0 otherwise. 
Specifically, for each measurement cell, we compute its average RSS value across all RSS maps generated at different heights, independently for each AP. 
We then bin the average RSS values of the measurement cells, with the bin size set to the standard deviation of the AP's RSS offset determined in \Cref{sss:sionna_calibration}. 
This allows us to account for cases in which two spatially close locations exhibit noticeable differences in RSS due to noise. 
\Cref{fig:rss_bins_ap_002} illustrates the resulting binned RSS heatmap for AP\_002.
To obtain the binary matrix for each AP, we determine whether the measured RSS value falls within the binned interval associated with each measurement cell and set the corresponding element of the binary matrix accordingly. 
For example, \Cref{fig:rss-select-bin-ap_002} shows the measurement cells whose average RSS values for AP\_002 fall between $-78.8$ and $-68.9$ dBm, corresponding to a reference RSS value of $-75.0$ dBm measured at the location marked by the red star.
\Cref{fig:ap_002_bin_image} shows the resulting representation of this sample according to AP\_002. This representation is used as input to our proposed localization model and two of our baseline methods that we will describe later.

\paragraph{Multivalued Matrix}
\label{sss:data_gen_method_heat_map}
The third input representation is a multivalued matrix for each AP, where each element corresponds to a measurement cell and its value captures the similarity between the cell's average RSS value and the measured RSS value. 
Specifically, for each measurement cell, we compute its average RSS value independently for each AP, following the same approach described in \Cref{sss:data_gen_method_rss_bin_maps}.
However, instead of using a binning strategy, we assign each cell a weight between 0 and 1 based on the similarity between its average RSS value and the measured RSS value. 
This weight can be interpreted as the brightness (color) of the corresponding pixel in the multivalued ``grayscale'' (colored) image, with higher weights indicating greater similarity between the two RSS values. 
Specifically, the weight assigned to each measurement cell is given by the Gaussian density function in \Cref{eq:normalized_gaussian_function}, where the mean is the measured RSS value at the reference location and the standard deviation is taken from the calibration process described in \Cref{sss:sionna_calibration}.
\begin{equation}
    g(x) = \frac{1}{\sigma\sqrt{2\pi}} \exp\left( -\frac{(x - \mu)^2}{2\sigma^2} \right)
    \label{eq:normalized_gaussian_function}
\end{equation}

\begin{figure}[h!]
    \centering
    \includeinkscape[width=0.4\textwidth]{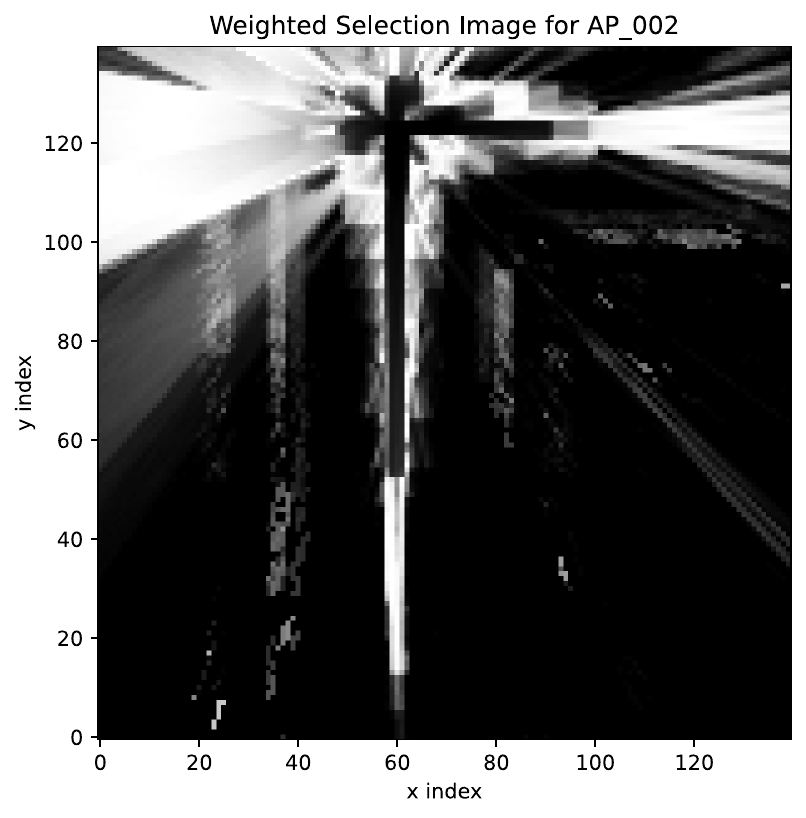}
    \caption{An example of a mutlivalued matrix, represented as a grayscale image constructed from RSS values for AP\_002}
    \label{fig:ap_005_gaussian_image}
\end{figure}

\Cref{fig:ap_005_gaussian_image} shows an example representation for AP\_005, in which the selection threshold is set to three times the standard deviation for AP\_002 determined during the calibration process.
This representation is used as input to our proposed localization model, with the goal of further improving the localization accuracy. 
\noindent

\subsection{Frequency Fusion}
Modern Wi-Fi technology supports the use of several frequency bands in order to improve performance and connectivity in radio-congested environments.
The most recent standards permit the 2.4, 5, and 6 GHz bands to service wireless devices.
Since frequency affects radio propagation within a physical environment, and since Sionna RT can produce radio maps across different frequencies with very little additional effort, we choose to consider the additional case of fusing data from multiple Wi-Fi bands in order to improve localization performance.
There are several ways in which measurements from different frequencies can be combined; 
the specific fusion method employed by each localization technique, together with the rationale behind it, is described in the corresponding methodology section.
Broadly speaking, two categories of fusion are explored: \emph{upstream fusion}, in which the readings from different frequencies are combined before being passed to the localization method, and \emph{downstream fusion}, in which separate location estimates are produced for each frequency and the resulting estimates are then combined into a single estimate.

\subsection{Localization Methodology}
\label{sse:localization_methods}
\begin{figure*}[h!]
    \centering
    \includegraphics[page=1,width=0.8\textwidth]{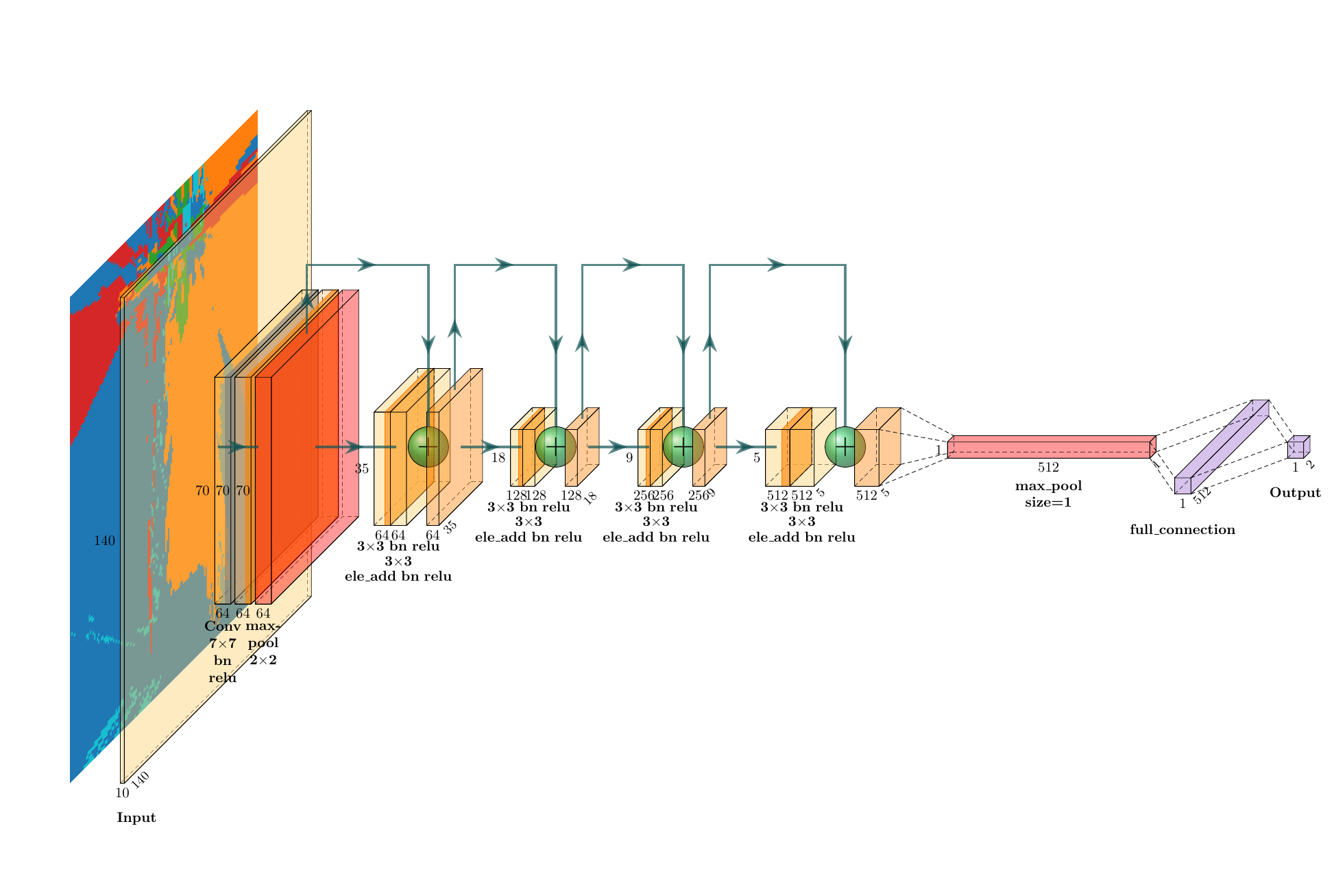}
    \caption{Model Architecture}
    \label{fig:ml_arch}
\end{figure*}

With the calibrated synthetic data in hand, we now turn to the localization method that consumes this data to estimate the user's location.
Our proposed localization method uses an architecture based on the ResNet-18 model, an 18-layer convolutional neural network~\cite{he_deep_2015}.
The model can take either binary or multivalued data as input and produce a location estimate.
Since each AP yields its own matrix, we stack the per-AP matrices along the channel dimension to form a single tensor.
This tensor has $N$ channels when only one frequency band is used, where $N$ is the number of APs, and $2N$ channels when both 2.4 and 5 GHz bands are used for upstream fusion since each band contributes its own matrix per AP.
The resulting multi-channel input is then processed by a ResNet model analogously to a conventional color image, with the network producing the predicted $x$ and $y$ coordinates of the user's location as output.
The deep learning architecture itself is based on ResNet-18, modified to accept $N$ input channels for single band localization or $2N$ channels for dual band localization.
Additionally, we replace the average pool near the end of the network with a max pool, and modify the final fully connected layer to predict the $x$ and $y$ coordinates of the user's location, instead of the 1,000 logits used for classification in ResNet. 
The architecture of the neural network model is shown in \Cref{fig:ml_arch}.

The ResNet model is implemented in PyTorch and trained for up to 100 epochs using mean squared error (MSE) loss.
The Adam optimizer is used with a learning rate of $0.0003$.
During training, $10\%$ of the training set is held as a validation set, and the validation loss is computed at the end of every epoch.
Early stopping is employed; training is terminated when the validation loss fails to improve for 20 consecutive epochs.
Once training has finished, the model weights corresponding to the lowest validation loss are loaded and evaluated on the test set.
Note that we do not train the model exclusively on real data, as our real RSS fingerprints (which are fewer than 50) are insufficient to train and evaluate the proposed ResNet model. 
Furthermore, to be more comparable to our baseline experiments, we also trained the model using a \emph{real-life seeded dataset} that adds a little bit of real-life data into the training set to be comparable to the related works, details of this is discussed in \Cref{sss:dataset_description}.
We repeat this experiment for both datasets 10 times with different random seeds for model training.

In \Cref{sss:evaluation}, we evaluated two different versions of this model: one that processes binary matrix to estimate the location and one that processes the multivalued matrix for the same task.

\section{Experimental Setup}
\label{sec:setup}
In this section, we describe ground truth data collection, our Sionna RT ray-tracing setup, synthetic data generation process, and finally the baseline localization methods. 
Our experimental environment was the first floor of Athabasca Hall (ATH) on the University of Alberta campus. The first floor of ATH  contains 37 offices, 4 bathrooms, 1 lounge, and 1 hallway divided into 6 segments.
This yields a total of 48 ``zones'' as defined by the BIM file of ATH.
Most zones are rectangular, with the exception of some hallway segments that take a ``L'' or ``T'' shape.
The origin of the coordinate system is taken to be the exterior point at the south west corner of ATH. 
This choice of origin simplifies the labelling of AP locations and Wi-Fi soundings. 

\subsection{Ground Truth Data Collection}
We collect ground truth RSS measurements within the building of interest.
These measurements serve two purposes: calibrating the material parameters used by Sionna RT's as outlined in Stage I in \Cref{sss:sionna_calibration}, and correcting for residual errors that Sionna RT does not model as outlined in Stage III in \Cref{sss:sionna_calibration}.
At each of the 34 measurement locations, 15 seconds of Wi-Fi measurement readings provided by the Android API were recorded using a Samsung Galaxy A15 5G smartphone held in a fixed position and orientation at waist height.
These readings contain the Wi-Fi AP's MAC address (BSSID), frequency, and signal strength of the received beacon frames sent by the APs within range.
Measurements are taken in office rooms to capture a diverse set of RF interactions with the various materials and objects that are typically found in such spaces.
Once all the measurements have been collected, we averaged the RSS values for each AP at every location, and further separated the readings by frequency band; that is, the 2.4 and 5 GHz measurements are averaged independently for each AP.

\subsection{Sionna RT Ray-tracing Setup}
\begin{figure}[ht]
    \centering
    \includegraphics[width=0.20\textwidth, angle=90]{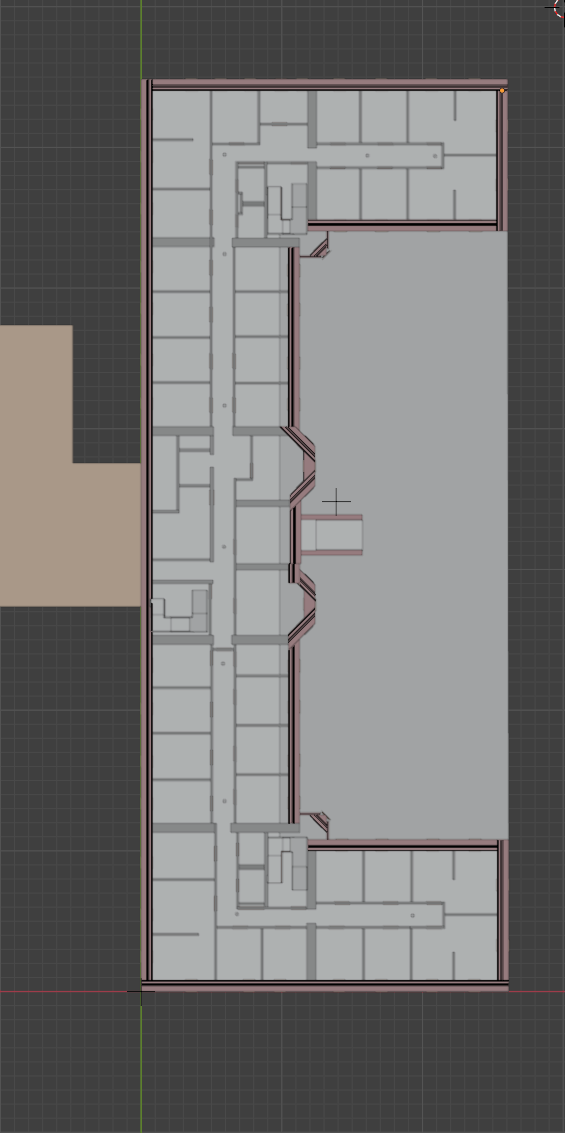}
    \caption{Overhead view of Level 1 of ATH. The South facing wall is to the right.}
    \label{fig:ath_floor_top_view}
\end{figure}

ATH is a 3 storey campus building with a semi-basement whose external walls are made of brick and interior walls made of drywall. 
This building was selected because we had access to a reasonably accurate BIM model and could readily collect ground-truth data within the offices and hallway.
To save time and reduce complexity, only the semi-basement (level-1) of ATH is modelled, although the same procedure can be extended to the other floors.
\Cref{fig:ath_floor_top_view} shows the overhead view of level-1 of ATH with the ceiling and the floors above removed for ease of viewing.
However, for ray tracing purposes, all floors and structures above level-1 are kept in the scene so that they continue to interact with the rays, mirroring the way a real radio signal would interact with the entire building.

\begin{figure}[h!]
  \centering
  \begin{subfigure}[h!]{0.4\textwidth}
    \includegraphics[width=\textwidth]{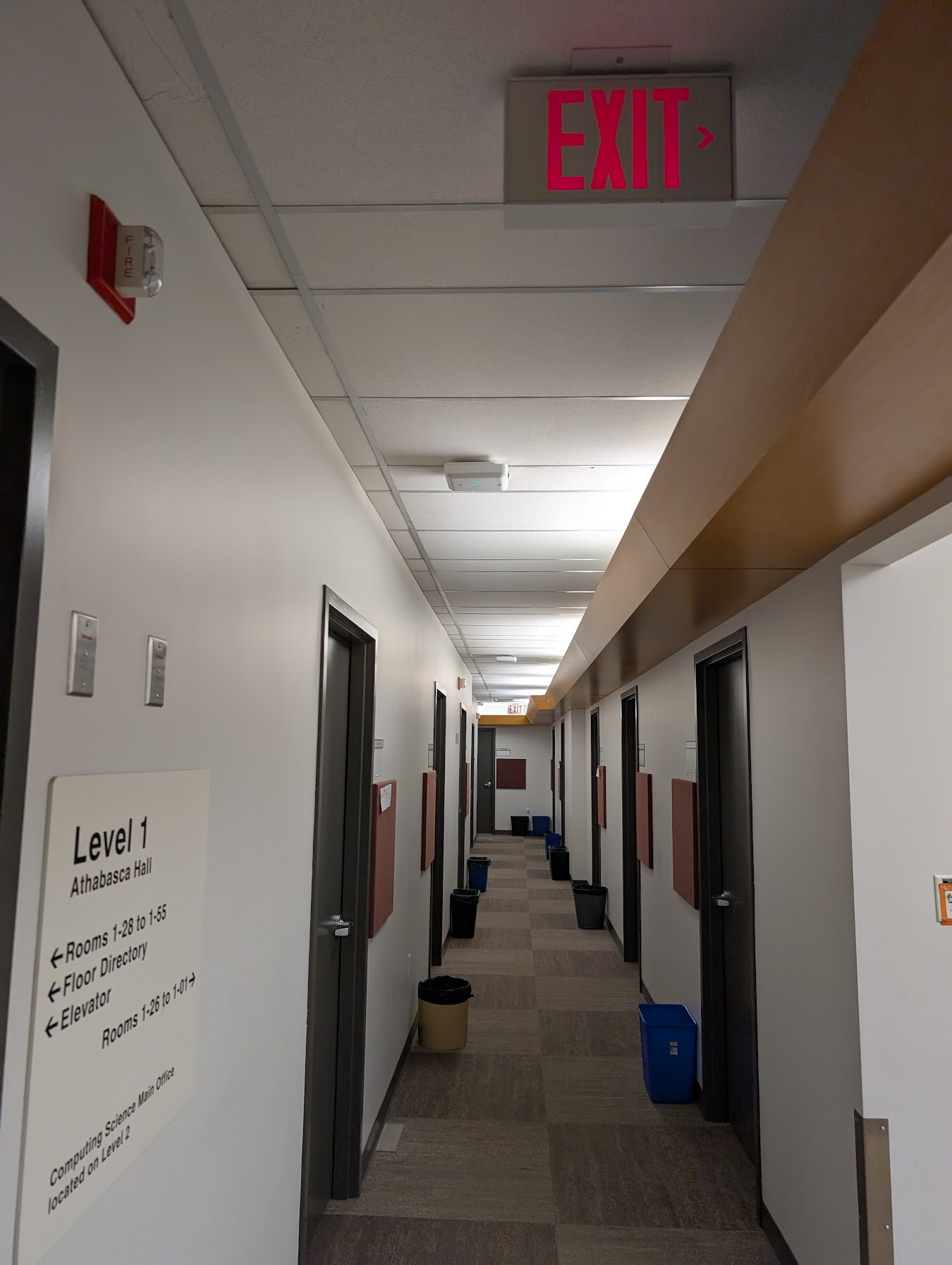}
    \caption{A long hallway inside ATH}
  \end{subfigure}
  \begin{subfigure}[h!]{0.4\textwidth}
    \includegraphics[width=\textwidth]{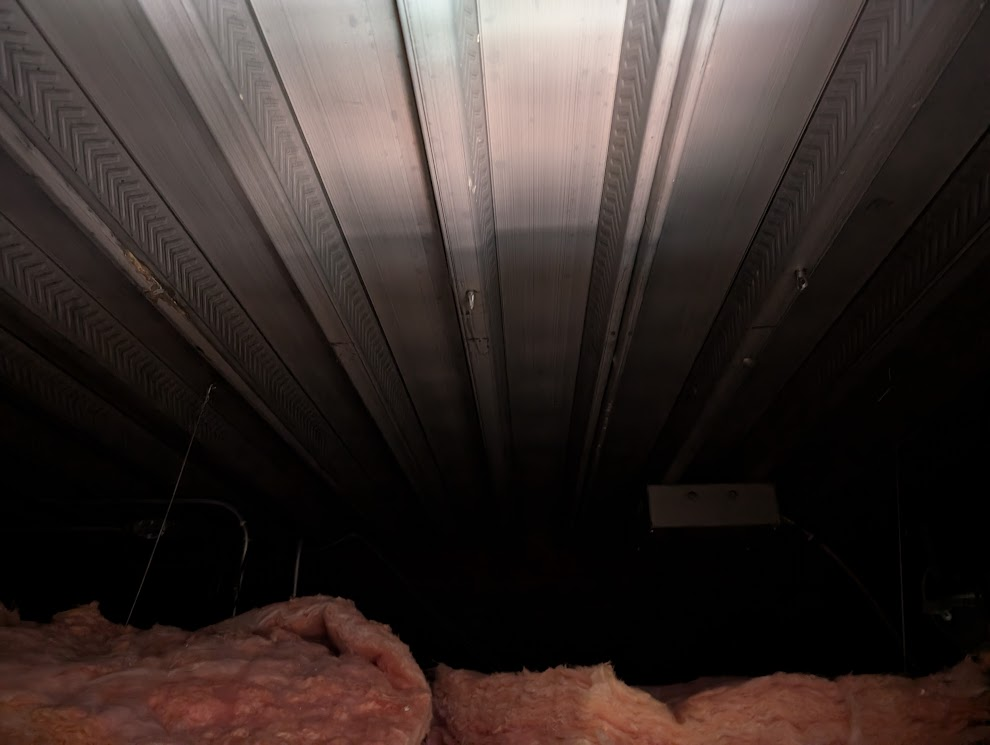}
    \caption{The area above the drop ceiling of ATH}
  \end{subfigure}
  \caption{Inside photos of ATH}
  \label{fig:inside_outside_photo_ATH}
\end{figure}

Before running the RF ray tracer in Sionna RT, we assigned a material type to each object that makes up the virtual environment.
The material types were determined through visual and tactile observation of the corresponding real-world environment.
\Cref{fig:inside_outside_photo_ATH} shows the interior of ATH; the hallways are predominantly made up of wooden doors, metal door frames, and drywall.
The ceiling is composed of ceiling board, with a metal sheet placed above it for fireproofing.
These material assignments were made in Blender and then exported as a Mitsuba scene file for use in Sionna RT.
\Cref{tab:best_material_thickness_params} reports the thickness of each material estimated using the Bayesian optimization method described in \Cref{sss:sionna_calibration}.
Note that, in some cases, a material represents a composite structure comprising of multiple components, such as drywall, insulation, and wall studs, which are lumped together and approximated as being a single materialwith a thickness determined by the Bayesian optimization method. 
For example, although a wall may consist of drywall, insulation, and wall studs, we can assign only a single material type (e.g., plasterboard) to the entire wall in the model. 
Consequently, the estimated thickness may be greater than the actual thickness of the assigned material within the composite structure.

\begin{table}[h!]
  \centering
  \begin{tabular}{lc}
    \hline
    \textbf{Material} & \textbf{Estimated Thickness (m)} \\
    \hline
    Material 1 & 0.5399 \\
    Material 2 & 0.0155 \\
    Material 3 & 0.4441 \\
    Material 4 & 0.8765 \\
    Material 5 & 0.8214 \\
    Material 6 & 0.1089 \\
    Material 7 & 0.9680 \\
    \hline
  \end{tabular}
  \caption{Material thickness used for RF ray tracing.}
  \label{tab:best_material_thickness_params}
\end{table}

There are 10 Wi-Fi APs on level-1 of ATH, whose real-life positions were determined by visually matching their physical locations to known reference points within the virtual environment.
These positions were then used to place the radio transmitters that Sionna RT employs to generate the radio map of the environment.
All transmitters share the same omnidirectional antenna configuration.
Each transmitter emits Wi-Fi signals on both the 2.4 and 5 GHz bands; the two bands are computed separately, since they require different transmitter parameters such as transmit power and carrier frequency.
We assume that each transmitter uses a single antenna with a common configuration shared between the 2.4 and 5 GHz bands.
\Cref{tab:transmit_power_for_APs} lists the transmit power, in dBm, used for each AP and frequency; 
these values were obtained by accessing Cisco Spaces Location API used by the building owners of ATH.

\begin{table}[h!]
  \centering
  \begin{tabular}{l|cc}
    \hline
    \textbf{AP Name} & \textbf{5.0 GHz (dBm)} & \textbf{2.4 GHz (dBm)} \\
    \hline
    AP\_000 & 17.0 & 5.0  \\
    AP\_001 & 17.0 & 11.0 \\
    AP\_002 & 14.0 & 11.0 \\
    AP\_003 & 14.0 & 5.0  \\
    AP\_004 & 11.0 & 5.0  \\
    AP\_005 & 14.0 & 5.0  \\
    AP\_006 & 14.0 & 8.0  \\
    AP\_007 & 8.0  & 8.0  \\
    AP\_008 & 17.0 & 8.0  \\
    AP\_009 & 20.0 & 11.0 \\ \hline
  \end{tabular}
  \caption{Transmit power for each AP for both Wi-Fi frequency bands}
  \label{tab:transmit_power_for_APs}
\end{table}

For our RF ray tracer, we enabled all ray interactions simulated by Sionna RT to obtain the most realistic RF simulation of the environment as possible.
The maximum interaction depth is set to 10, with a measurement cell size of 0.5 by 0.5 metres.
We set the samples per transmitter to 100,000,000 to capture as many material interactions as possible for each measurement cell.
The radio map itself spans 70 by 70 meters, which is sufficient to cover, and exceed, the entire area of the building floor plan.
Our RF ray-tracing and localization experiments were conducted on a server equipped with an Intel Core I9-9940X, 128 GB of RAM, and a Nvidia A5000 GPU. 

\subsection{Synthetic Dataset Generation}
\label{sse:synthetic_dataset_generation}
Since both a BIM model of ATH and a set of real Wi-Fi recordings within the same building are available to us, we were able to generate synthetic data for the first floor of ATH.
Although recordings from other floors of ATH are also available, we restrict our attention to a single floor to limit the scope and complexity of multi-floor localization.

When generating a radio map in Sionna, the user can specify the size, shape, and orientation of the measurement plane, as well as the size of the measurement cells.
These parameters control both the number of measurements that populate the dataset and the precision of those measurements.
Notice that, Sionna produces measurements for locations that are implausible locations for a device, e.g., between the walls of adjacent rooms, inside the brick walls forming the exterior of the building, etc. Given that we have access to the building geometry, those implausible locations are excluded from consideration.

Due to the deterministic nature of the RF solver in Sionna RT, repeatedly running the radio map solver on the same measurement plane (i.e. height) yields identical results. To introduce a degree of uncertainty for the RSS measurements at a location synthesized by Sionna RT, we generate 11 separate deterministic radio maps for each AP. At each point on the x,y plane, 11 measurements are synthesized differing in the elevation. The 11 elevation values correspond to 1 centimetre intervals between 1.2 (inclusive) and 1.3 (inclusive) meters, plus an additional midpoint at 1.25 meters.  The choice of the elevation is informed by the typical height where a user might hold a smartphone. The elevation-based diversity of  RSS introduces a degree of uncertainty of the RSS measurement taken at an x,y location despite Sionna RT's deterministic behavior.

\subsubsection{Construction of Training, Validation, and Test Sets}
\label{sss:dataset_description}
To investigate whether synthetic data alone are sufficient for accurate indoor localization, we compare it against an augmented dataset produced by seeding 20\% of the ground truth data into the synthetic dataset, referred to as the \emph{real-life seeded dataset}.
This is done to more closely mirror the dataset properties of the related work against which we compare, where augmentation is built around real measurements; in our case, by contrast, the synthetic dataset is itself generated from a simulator that has been calibrated against real sounding data.
The test set is the ground truth measurement gathered from the real environment that was used to calibrate our Sionna RT model. 

For our localization method and any baseline that incorporates a machine learning model, training is repeated 10 times on both the synthetic-only and real-seeded datasets.
At each repetition, a different seed is used to randomize the ordering of both datasets, as well as the random selection of the 20\% of real data used to augment the real-seeded dataset.
The 20\% selected for training is then removed from the test set in order to avoid leakage between training and evaluation.
For both datasets, we split 10\% of our training set into a validation set used during training.

\subsection{Baseline Localization Methods}
We consider four baseline localization methods: two non-learning-based methods developed in this work and two machine-learning-based methods adopted from related work. The two non-learning-based methods estimate a user's location using, respectively, the intersection of convex hulls and density-based clustering.

The two learning-based baseline methods from related work use models that we train and evaluate on the same augmented RSS dataset used for our proposed localization method and the other two baselines. 
Because these models were not originally designed to consume the two image-like representations, we use the RSS vector, the first representation introduced in \Cref{sss:transformations}, as their input.
Additionally, we only used the 2.4 GHz bands for these learning-based baseline methods, since those experiments used datasets that were exclusively on the 2.4 GHz band.
In contrast, the two non-learning-based baseline methods take as input the binary  matrices, the second representation introduced in \Cref{sss:transformations}.
This section describes the design and implementation of these baseline methods, which we use to evaluate the performance of our proposed localization method.

\subsubsection{Convex Hull}
\label{sss:convex_hull}
This baseline method takes as input the binary matrices for all reachable APs, which are generated using the transformation described in \Cref{sss:data_gen_method_rss_bin_maps}.
For each binary matrix, which can be viewed as a set of selected measurement cells, we compute the convex hull enclosing these cells. Since a convex hull is the smallest convex shape that contains a given set of points, it provides a natural starting point for a geometric localization technique. Once the convex hulls are computed for all reachable APs, we take their intersection and use the centroid of the intersection as the estimated user location. The intersection contains measurement cells that are selected by (voted for) all reachable APs and therefore represents a form of consensus among the APs.

\begin{figure}[t!]
    \centering
    \begin{subfigure}[t]{0.35\textwidth}
        \centering
        \includeinkscape[width=0.95\textwidth]{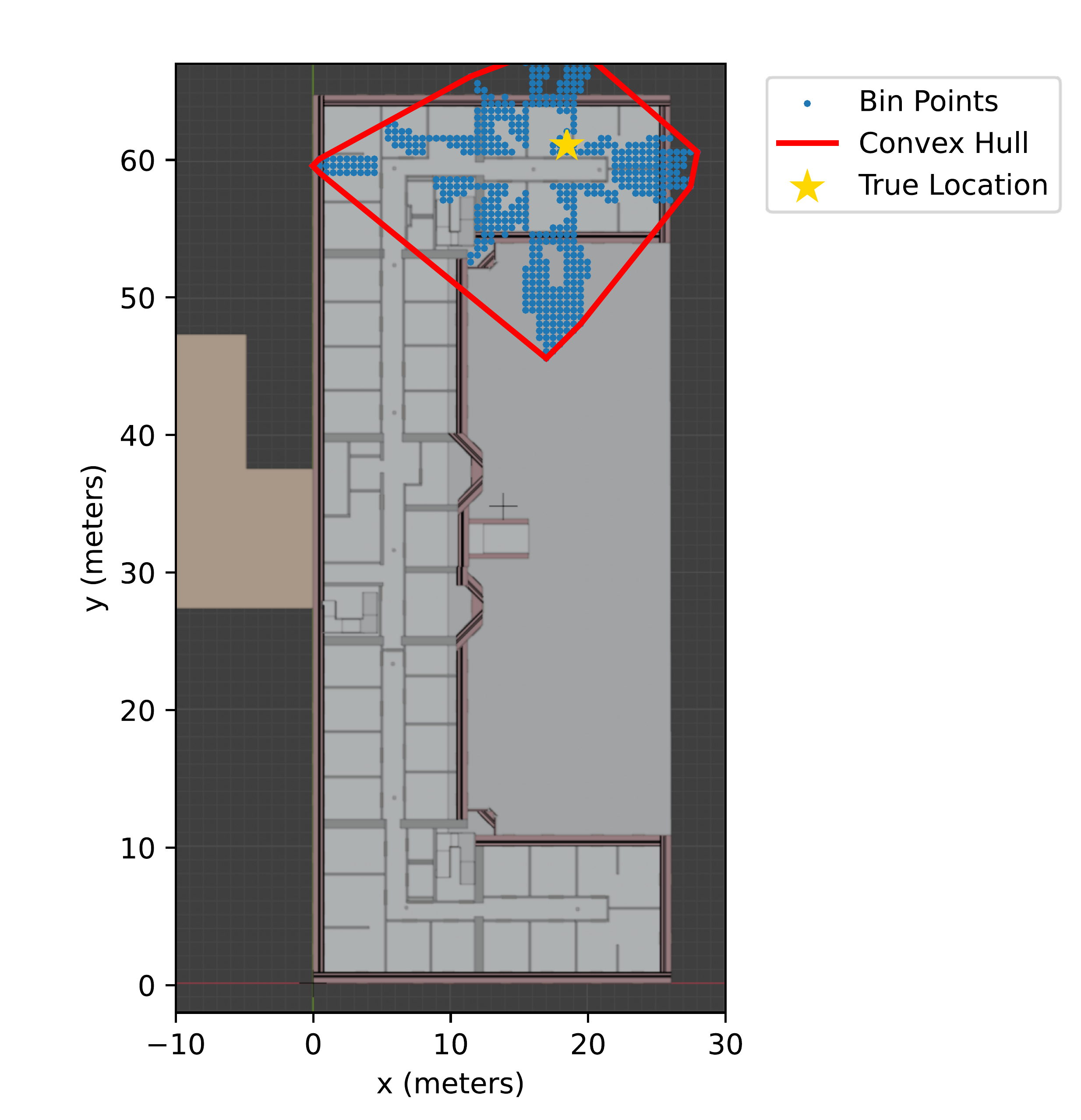}
        \caption{Convex hull of the selected bin for AP\_001}
        \label{fig:convex_hull_ap_001}
    \end{subfigure}
    \hfill
    \begin{subfigure}[t]{0.31\textwidth}
        \centering
        \includeinkscape[width=0.95\textwidth]{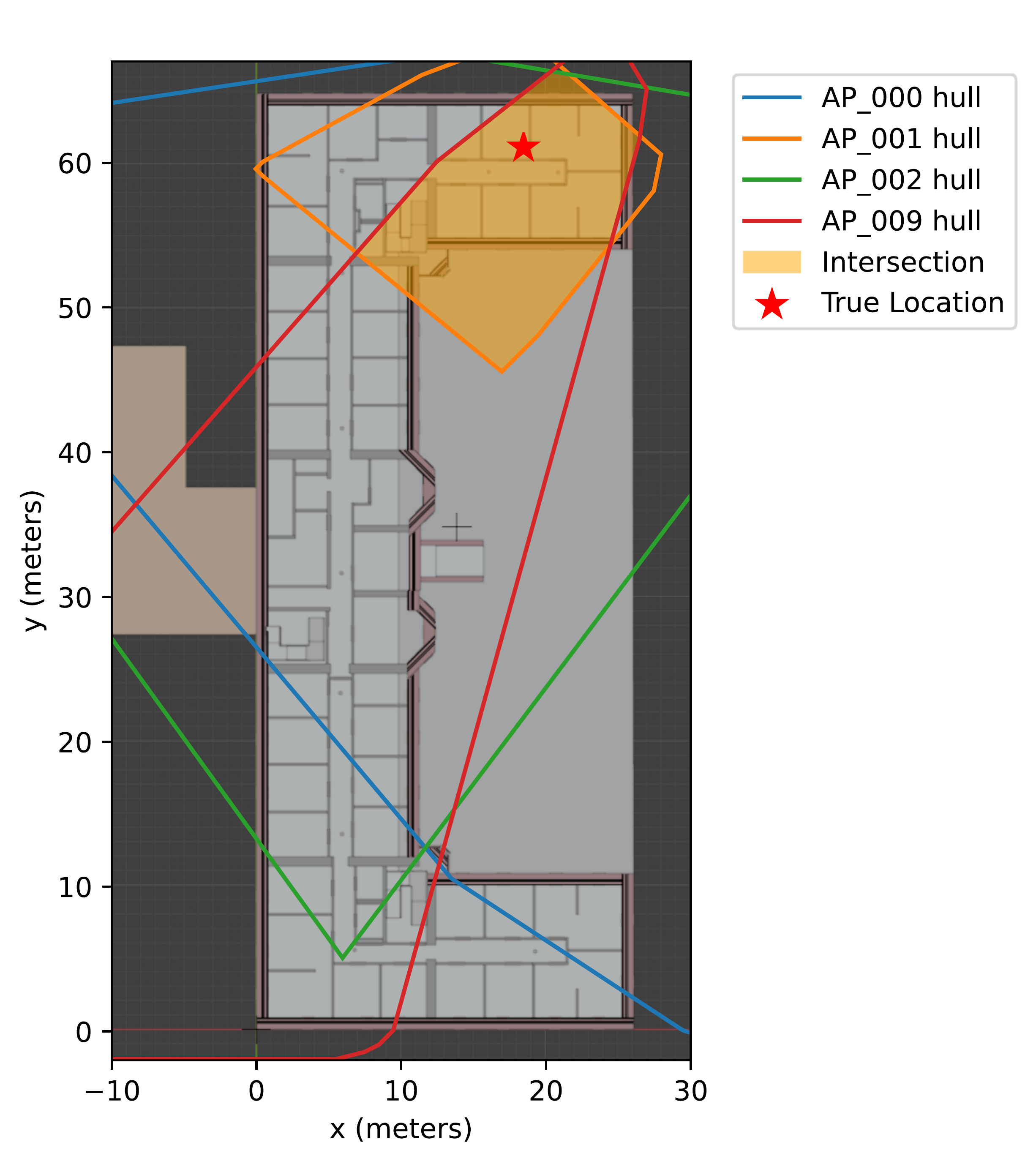}
        \caption{Intersection of all convex hulls}
        \label{fig:intersect_all_hull}
    \end{subfigure}
    \hfill
    \begin{subfigure}[t]{0.31\textwidth}
        \centering
        \includeinkscape[width=0.95\textwidth]{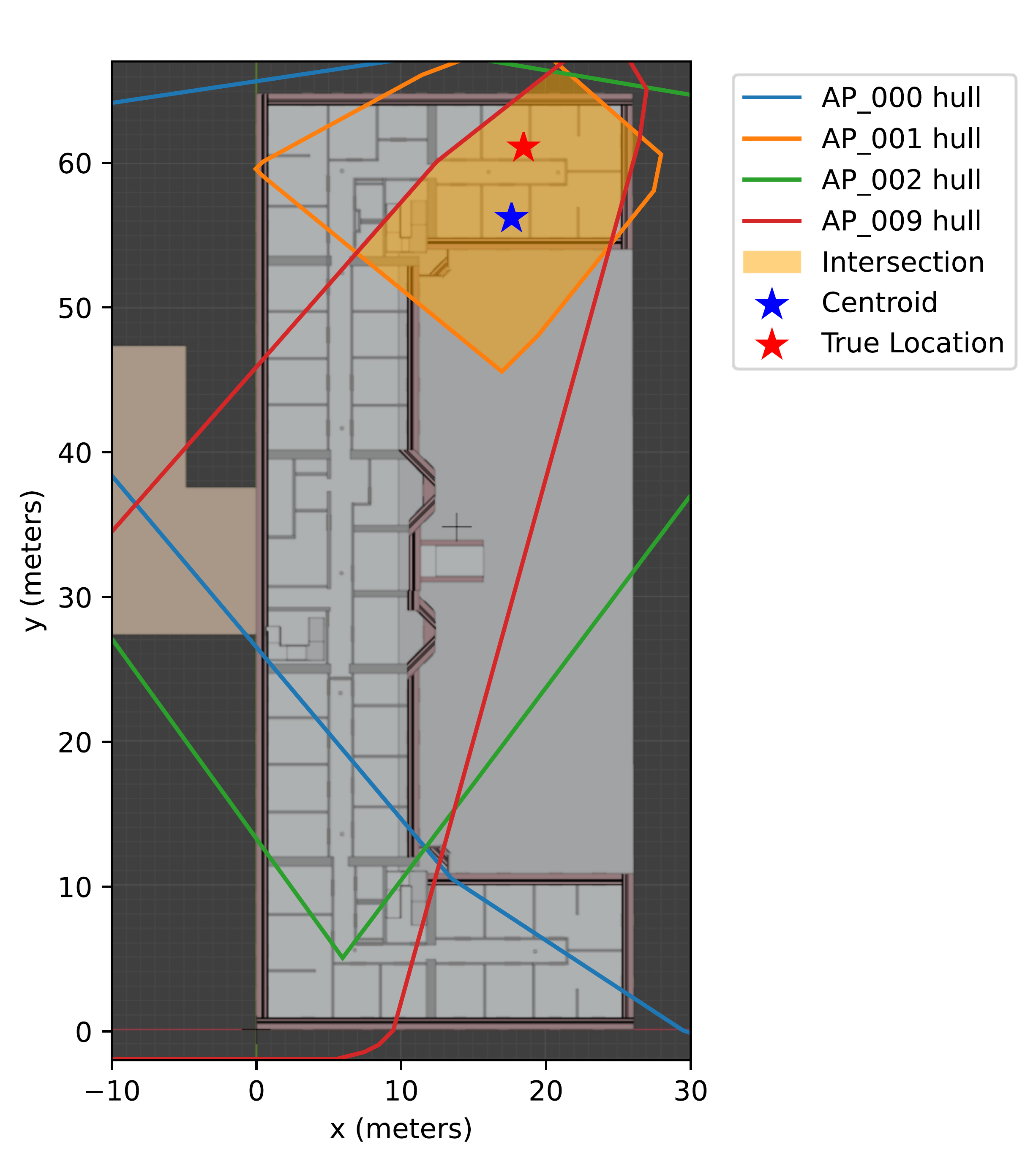}
        \caption{Location estimate based on the centroid of the resulting intersection}
        \label{fig:centroid_intersect_all_hull}
    \end{subfigure}
    \caption{Illustration of the convex hull-based localization baseline}
    \label{fig:convex-hull-method}
\end{figure}

If the intersection is empty, we reduce the required number of votes by one and recompute the intersection, repeating this process until a non-empty set of measurement cells is obtained. We select the smallest non-empty set because convex hulls tend to overestimate the region containing the user; thus, a smaller intersection corresponds to lower localization uncertainty. Intuitively, this approach identifies the smallest region that is supported by the largest possible number of reachable APs. The process is illustrated in \Cref{fig:convex-hull-method}.

\paragraph{Frequency Fusion:}
Downstream fusion combines location estimates from the two frequency bands by taking the midpoint between the location estimates produced independently for each band.
However, it may not improve localization performance, since each frequency is processed in isolation and no information is shared between the two until after a location has already been estimated.
An alternative is an \emph{upstream} fusion strategy that computes the intersection of all convex hulls produced by both frequencies and take the centroid of that combined intersection as the location estimate.
In this scheme, the two frequencies contribute jointly to the consensus region before any location decision is made.

\subsubsection{DBSCAN Clustering}
Similar to the previous baseline, this one also takes as input the binary matrices for all reachable APs.
It then applies DBSCAN, a density-based clustering method, to identify at least one spatially coherent group of measurement cells while suppressing outliers that would otherwise distort the location estimate.
The intersection of the clusters across all APs is then computed; if the intersection is empty, the algorithm proceeds in the same fall-back manner as in the convex hull method, reducing the number of required APs by one and retaining the largest non-empty intersection.
We select the largest intersection if we have multiple solutions to increase the probability of the clusters containing the true location. 
After computing the intersection of the clusters across all APs, a second invocation of DBSCAN finds the largest cluster in the intersection and returns its centroid as the estimated location.

\begin{figure}[t!]
    \centering
    \begin{subfigure}{0.45\linewidth}
        \includeinkscape[width=\linewidth]{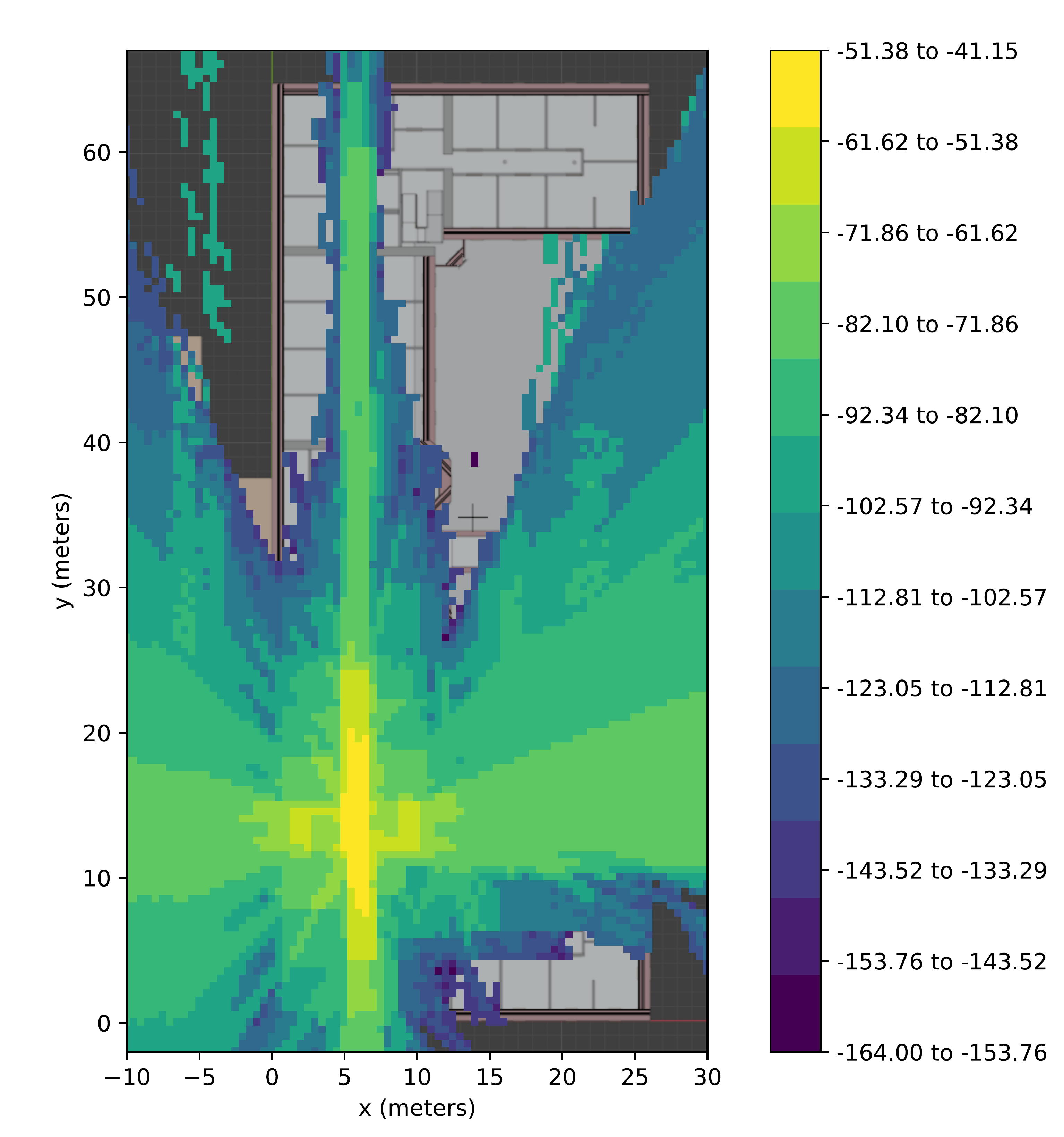}
        \caption{The bins for AP\_007 with a bin interval of 10.24 dBm}
        \label{fig:RSS_bins_for_ap_007}
    \end{subfigure}
    \hfill
    \begin{subfigure}{0.45\linewidth}
        \includeinkscape[width=\linewidth]{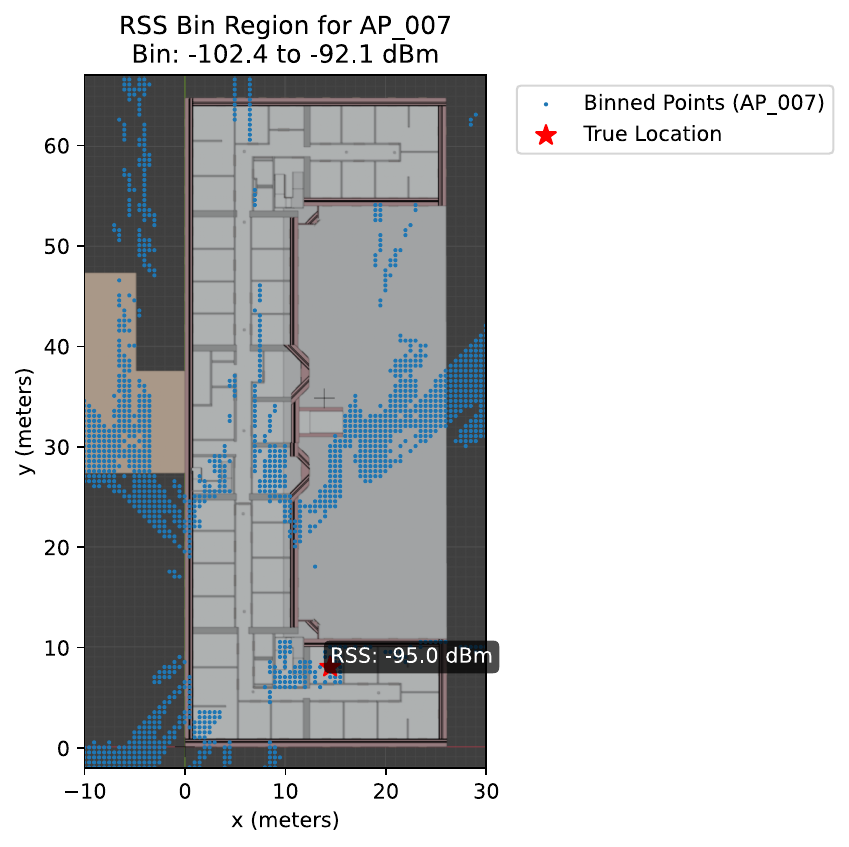}
        \caption{The select bin for AP\_007 using a reference value of -95.0 dBm}
        \label{fig:RSS_bin_selection_ap_007}
    \end{subfigure}
    \caption{The binning and bin selection process for AP\_007 for a reference value of -95.0 dBm}
    \label{fig:bin_and_bin_selection_for_ap_007}
\end{figure}

\begin{figure}[t!]
    \centering
    \begin{subfigure}[h]{0.45\linewidth}
        \centering
        \includeinkscape[width=\linewidth]{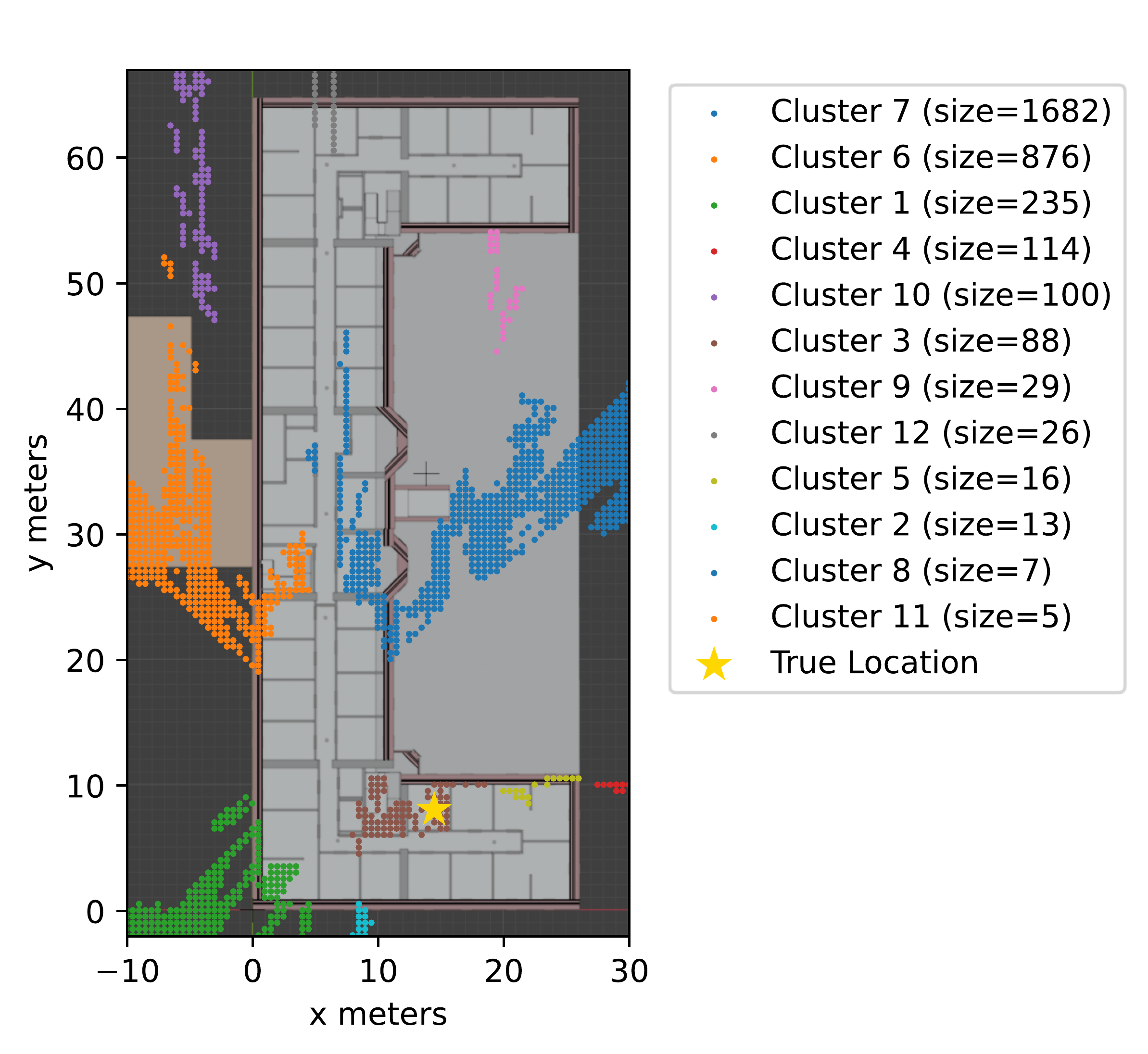}
        \caption{The clusters from the selected bin for AP\_007}
        \label{fig:dbscan_clustering_ap_007}
    \end{subfigure}
    \hfill
    \begin{subfigure}[h]{0.24\linewidth}
        \centering
        \includeinkscape[width=\linewidth]{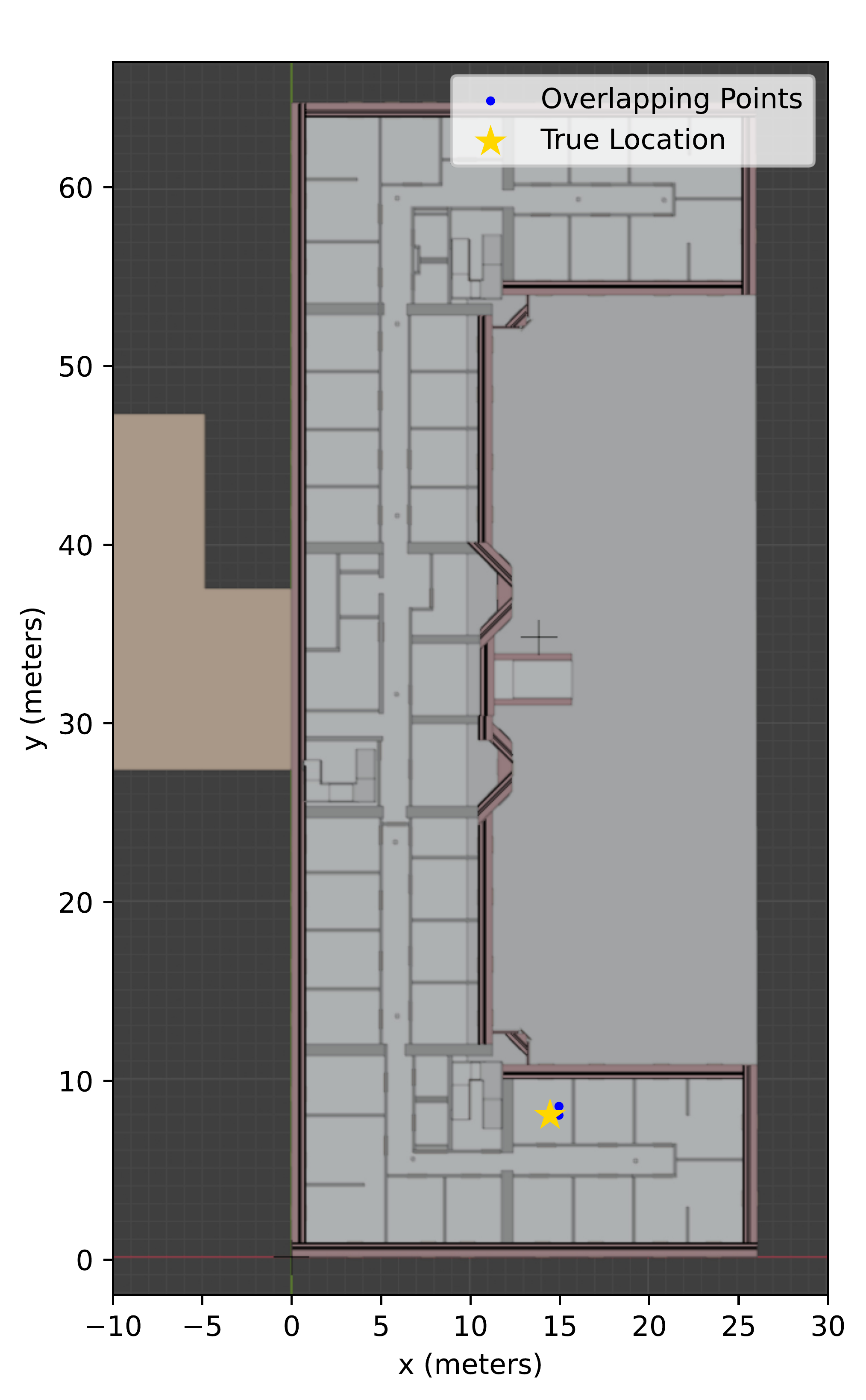}
        \caption{The intersection of all clusters across all APs}
        \label{fig:intersecting_all_dbscan_bins}
    \end{subfigure}
    \hfill
    \begin{subfigure}[h]{0.26\linewidth}
        \centering
        \includeinkscape[width=\linewidth]{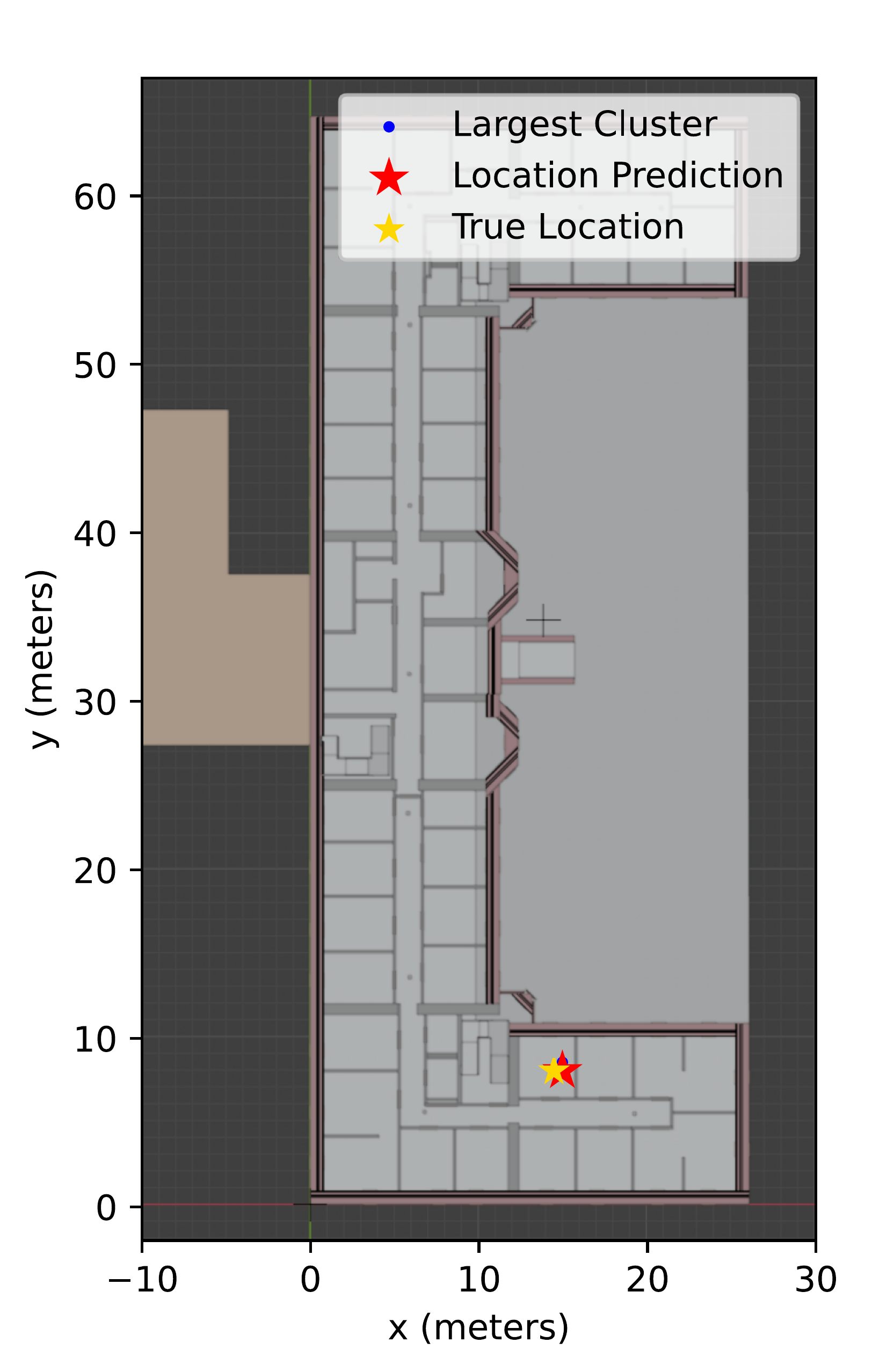}
        \caption{The largest cluster from all the intersecting bins}
        \label{fig:largest_dbscan_cluster_bins}
    \end{subfigure}
    \caption{Illustration of the DBSCAN-based localization baseline}
    \label{fig:DBSCAN}
\end{figure}

\Cref{fig:bin_and_bin_selection_for_ap_007} shows an example of the binary matrix/image containing a subset of measurement cells for AP\_007, given the reference RSS value of $-95.0$ dBm.
Within the set of measurement cells selected for each AP, DBSCAN is applied with $\varepsilon$ representing the maximum distance between two samples for one to be considered in the neighbourhood of the other set to $1.0$ and a minimum sample count of $5$ in order to identify the per-AP clusters.
\Cref{fig:DBSCAN} shows the 12 clusters found by DBSCAN for AP\_007 using different colors. 
A second invocation of DBSCAN finds the largest cluster in the intersection; 
for this stage, $\varepsilon$ is reduced to $0.5$ while the minimum sample count is left unchanged.
The lower $\varepsilon$ value at the intersection stage is chosen in order to mitigate the risk of forming spurious ``superclusters'' from nearby ``islands'' of clusters.
The estimated location which is the intersection of the identified clusters for all APs and the ground truth location are shown in the next subplots.

\paragraph{Frequency Fusion:}
The downstream fusion strategy takes the midpoint between the location estimates computed independently for each frequency band.
It suffers from the same limitation noted in \Cref{sss:convex_hull}: the two frequencies do not share any information until the very end of the pipeline.
To allow some information sharing prior to the final estimate, we introduce a refined downstream fusion strategy, named \textit{Downstream Overlap Fusion}, in which the largest clusters from the 2.4 and 5 GHz estimates are intersected to obtain a region of consensus, and the centroid of that consensus region is then taken as the estimated location.
The upstream fusion strategy pools together clusters from both bands across all APs to find the largest overlapping cluster; the rest of the localization procedure is the same as the single band method.

\subsubsection{RBF}
Our third localization baseline is based on the RBF network proposed in~\cite{laoudias_localization_2009}.
Since our Sionna-generated radio map is too dense for each $(x, y)$ location to serve as its own kernel centre, we aggregate the measurement cells to the nearest whole meter and use the corresponding average RSS values as kernel centres.
This aggregation yields 319 kernel centres on which the network is trained.
The radial and norm functions used by the network match those of the original RBF work, namely a Gaussian radial function and the Euclidean norm.
The $\beta$ parameter of the Gaussian radial function is computed from the heuristic given below, which is taken directly from the original RBF paper~\cite{laoudias_localization_2009}:
\begin{equation}
  \beta = \frac{1}{2d_{max}}.
  \label{eq:rbf_beta_calculation}
\end{equation}
Here $d_{max} = \text{max} || c_i - c_j||$ for $i,j = \{1, \dots, L\}$ with $L$ being the number of locations of all reference points and $c_i$ and $c_j$ being any two locations.
Since only real data was used to train the RBF network in the original study, we used the real-life seeded dataset outline in \Cref{sss:dataset_description} to more closely resemble the composition of the data set in the article. 
Training is carried out for 500 epochs using the Adam optimizer, with a learning rate of $0.001$ and MSE as the loss function; then the trained model is evaluated on the test set with the real-life seeded training data removed.
We train and independently evaluate the RBF network 10 times using different random seeds.

\subsubsection{extendGAN+}
Our last localization baseline is extendGAN+, proposed in~\cite{yean_extendgan_2023}. 
This method integrates data augmentation using WGAN-GP with localization using a ResNet-based architecture.
We follow the experimental setup of the original paper as closely as possible, including implementing their augmentation pipeline using our real-life RSS measurements. 
Since training the WGAN-GP model requires real data, we randomly partition our real RSS dataset such that 80\% of the real data (collected at 27 out of the 34 locations) is used to train the WGAN-GP augmentation model and the remaining 20\% is used to evaluate the localization model trained on the augmented data. We repeat this partitioning 10 times, as in our other experiments, to obtain broad test coverage across our real data.
The rest of the augmentation process remains unchanged.
The augmented training data used for localization consists of the original ground-truth measurements (used to train the WGAN-GP model), samples generated from Dirichlet upsampling, and synthetic data generated from the WGAN-GP model after filtering.
The test set used to evaluate the localization model averages the RSS measurements from all APs into a single record to remain consistent with the test set used in our other experiments. The size of the training set used to train their localization model ranges from 540 to 648 samples, depending on which 27 location measurements are selected for augmentation.
The test set consists of the 7 measurement locations that were not used for augmentation.

In extendGAN+, multiple localization models were considered, including a random forest, a deep neural network, and different ResNet-based architectures. We selected the best-performing model across all environments evaluated in the original work, which was a ResNet-18 modified to accept a single-channel input and produce two output features, one for each predicted coordinate, $x$ and $y$. 
The feature vector is of size $4 \times 4$, which is sufficient to encode all 10 APs; the remaining entries are zero-padded, and the input vector is normalized. 
It is important to note that the ResNet-18 model used in the extendGAN+ baseline has a few differences from the ResNet-18 model used in our work for localization. First, our input is an image-like matrix representation of possible locations rather than an RSS matrix representing measurements from all APs. Second, our model uses also frequency fusion that combines data from two frequency bands rather than the single 2.4 GHz band that was used in the original paper because that was the only frequency band available in their dataset.

To evaluate the localization model independently of their augmentation method, we use our augmented real-life seeded dataset to train the model with a data composition that more closely matches the dataset used in extendGAN+.
The seeded dataset is converted into a two-dimensional RSS feature vector matching the format produced by the extendGAN+ augmentation process.
We call this baseline ``extendGAN+ with ray-traced augmentation'' in \Cref{sss:evaluation} to highlight that it uses our ray-tracing-based augmentation method. 
To evaluate the complete augmentation and localization pipeline proposed in extendGAN+, we use their augmentation technique with our real-life measurements.
We call this baseline ``extendGAN+'' in \Cref{sss:evaluation}.
The Adam optimizer is used with a learning rate of $0.0001$, and training is run for up to 500 epochs with early stopping triggered when the validation loss, computed after every training epoch, has not improved for 25 consecutive epochs. These are the same training hyperparameters used in the original work, and we use them for both the ``extendGAN+ with ray-traced augmentation'' and ``extendGAN+'' baselines.
Once training has finished, the weights corresponding to the lowest validation loss are loaded and evaluated on the test set, with the real-life seeded training samples excluded from the test set. We train and independently evaluate the ResNet model 10 times using different random seeds.

\section{Results}
\label{sec:results}
In this section we first present the result of calibrating Sionna RT, then discuss the coverage of the synthetic dataset that we generated, and finally evaluate our localization model against the baselines.

\subsection{Verifying the Quality of Synthetic Data Generated by Sionna}
\label{sse:sionna_validation}
We compare real RSS fingerprints with the synthetic ones derived from Sionna RT to verify that our calibrated simulator can produce realistic RSS values.
\Cref{tab:2.4_5_validation_results} shows the mean absolute error after calibration for each AP.
We found that across 34 measurement locations, the average difference between the calibrated RSS value and the real RSS value is 4.970 dBm for all 10 APs on the 2.4 GHz band and 7.269 dBm for all 10 APs on the 5 GHz band.

\begin{table}
    \centering
    \begin{tabular}{lccccccccccc}
        \toprule
        Access Point& AP\_000 & AP\_001 & AP\_002 & AP\_003 & AP\_004 & AP\_005 & AP\_006 & AP\_007 & AP\_008 & AP\_009 \\
        \midrule
        \makecell{Mean Absolute\\Error (dBm)\\at 2.4 GHz}
        & 4.645 & 6.837 & 4.371 & 5.559 & 4.737 & 3.442 & 6.373 & 6.614 & 3.888 & 3.236 \\
        \makecell{Mean Absolute\\Error (dBm)\\at 5 GHz}
        & 11.899 & 9.958 & 7.642 & 8.015 & 4.804 & 7.421 & 5.559 & 7.035 & 3.927 & 6.426 \\
        \bottomrule
    \end{tabular}
    \caption{Calibrated mean absolute error for both 2.4 and 5 GHz frequencies}
    \label{tab:2.4_5_validation_results}
\end{table}

\begin{figure}[h!]
    \centering
    \includeinkscape[width=0.75\linewidth]{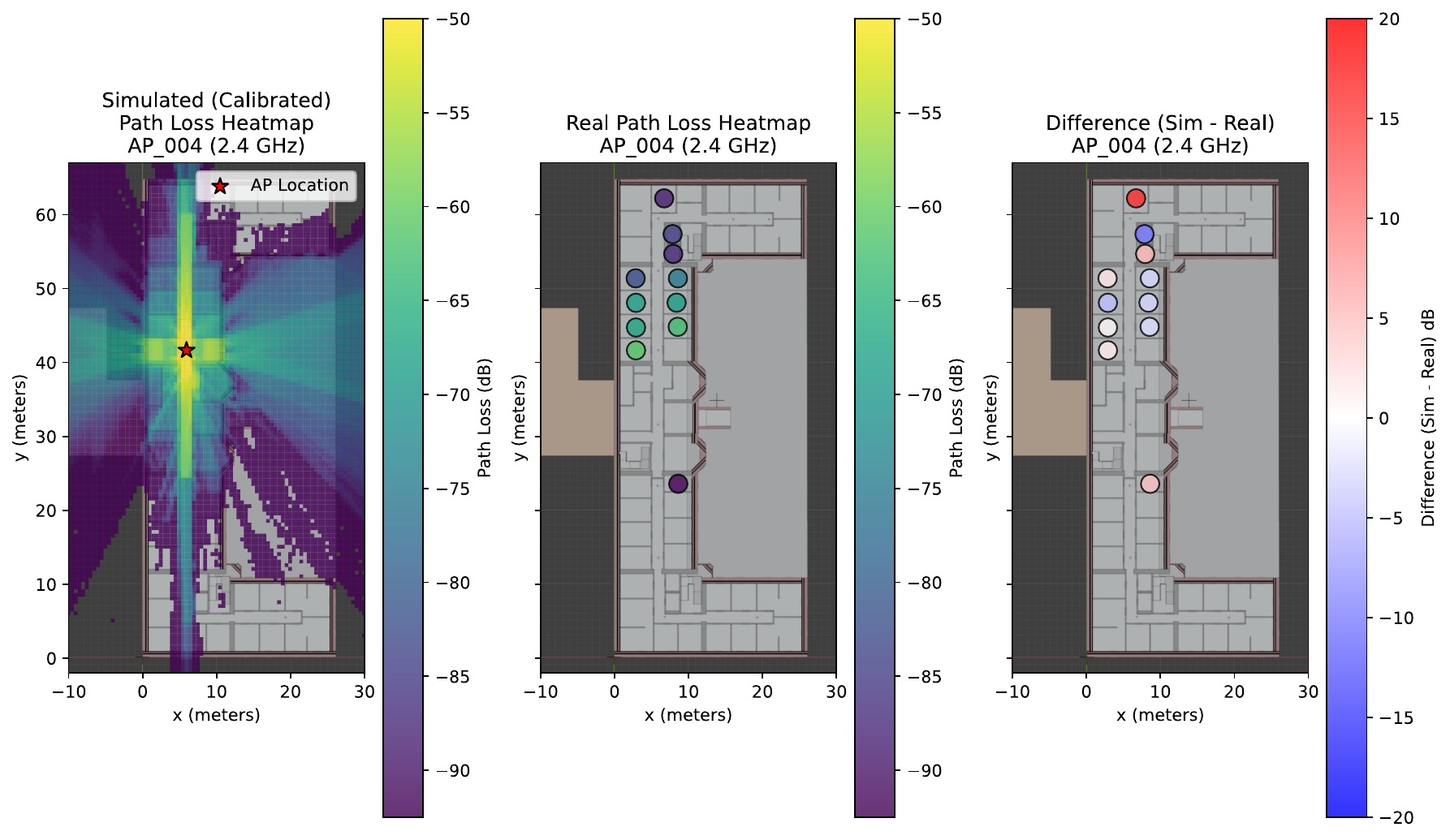}
    \caption{Path Loss Differences for AP\_004 between the Calibrated Simulated Values and Real Values for Wi-Fi (2.4 GHz)}
    \label{fig:wifi_2.4_sim_validation_results}
\end{figure}

\begin{figure}[h!]
    \centering
    \includeinkscape[width=0.75\linewidth]{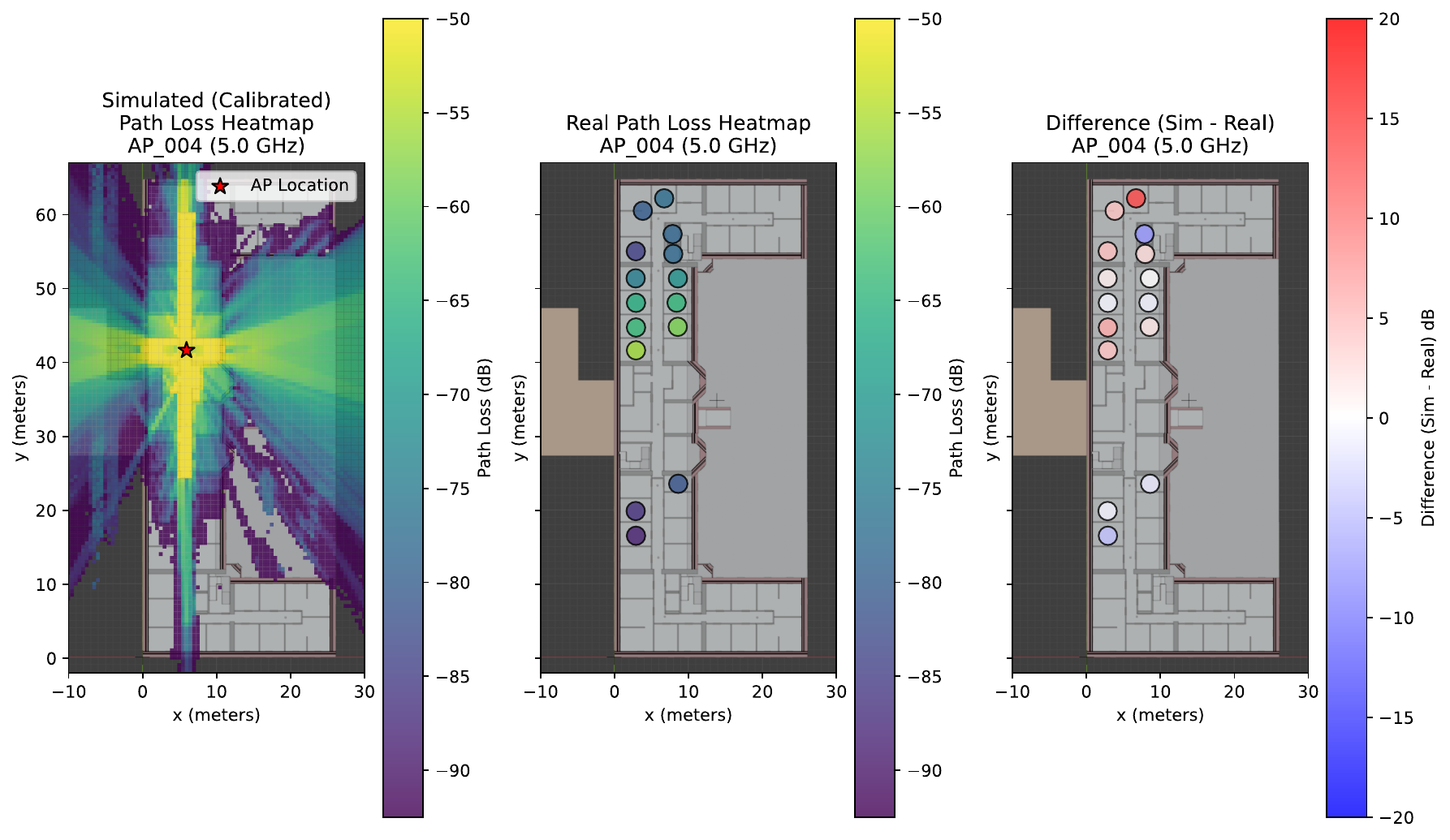}
    \caption{Path Loss Differences for AP\_004 between the Calibrated Simulated Values and Real Values for Wi-Fi (5 GHz)}
    \label{fig:wifi_5.0_sim_validation_results}
\end{figure}

By way of comparison, the global offset calibration method used in the material parameter search from \Cref{sss:sionna_calibration} achieved an RMSE of 6.83 dBm on the 5 GHz band; the gap between this figure and our per-AP offset results is small enough that the choice of per-AP versus global calibration is unlikely to drastically affect outcomes downstream.
\Cref{fig:wifi_2.4_sim_validation_results} and \Cref{fig:wifi_5.0_sim_validation_results} show the differences between the calibrated simulated values and the real values for AP\_004 on the 2.4 GHz and 5 GHz bands respectively.


\subsection{Coverage of the Synthetic Dataset}
\label{sse:synthetic_dataset_result}
\begin{figure}[h!]
    \centering
    \includeinkscape[width=0.30\textwidth]{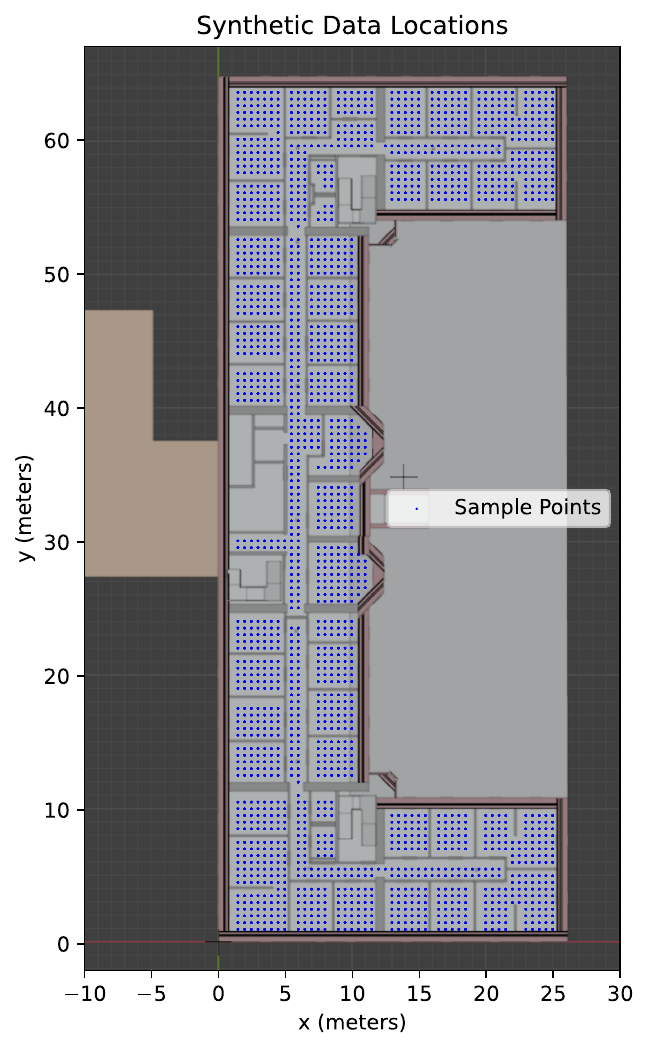}
    \caption{Measurement locations for our synthetic dataset}
    \label{fig:syth_data_locations}
\end{figure}

Each radio map produced for level-1 of ATH contains 19,600 measurement locations per AP, with a separate map generated for each of the 10 APs.
Thus, stacking the 11 height layers brings the total to 215,600 samples across the three spatial dimensions. 
After pruning 191,895 measurement cells whose footprint involves implausible locations (as described in  \Cref{sse:synthetic_dataset_generation}), we retain 2,155 measurement locations, corresponding to 23,705 synthetic samples in total.
The locations of the measurement points used in our dataset are shown in \Cref{fig:syth_data_locations}. 
As shown in the figure, the measurement locations provide good spatial coverage throughout the indoor environment.

\subsection{Evaluation of Our Localization Method}
\label{sss:evaluation}
For a fair comparison, we evaluate all localization methods, including the proposed method and the four baselines, using the dataset described in \Cref{sse:synthetic_dataset_generation}.
We use as evaluation metric, the mean Euclidean distance between the estimated location and the location at which the corresponding ground truth measurement was taken.
The complete set of results is reported in \Cref{tab:full_results}, which lists the mean localization error of each method together with the standard deviation over individual experiments, broken out by frequency band and by training-data regime.
 
Across all experiments, the proposed localization method with upstream fusion and multivalued matrix as input performs best, achieving a mean localization error of 3.05 meters on the real-data test set when trained on the synthetic-only dataset.
It outperforms the best baseline method, namely extendGAN+, whose mean error on the same test set was 4.59 meters, \emph{i.e.} our localization method reduces the error by 1.54 meters compared to the baseline's best method.
Notably, the mean standard deviation of the best-performing variant of our localization method is also lower than all baselines, indicating our location estimates are more stable compared to the best related works.

For single frequency performance, the 2.4 GHz band outperformed the 5.0 GHz band across all of our experiments.
The improved performance is most likely caused by the better material penetrating properties of the 2.4 GHz band compared to the higher 5.0 GHz band, thus giving our model more accurate RSS measurements.
This is further seen by the lower mean absolute error in the 2.4 GHz band compared to the 5.0 GHz band as shown by \Cref{tab:2.4_5_validation_results} where across most APs the 2.4 GHz band has a better mean absolute error by a few dBms on average compared to their 5.0 GHz counterpart.
The best performing method using a single frequency band (2.4 GHz) was the proposed localization method with multivalued matrix as input, achieving a mean localization error of 3.56 meters and 3.55 meters when using the synthetic only and real-life seeded datasets respectively.

In both the synthetic only and real-life seeded datasets, the proposed localization method with mutlivalued matrices as input produced the best and second best results across all experiments. 
This suggests that giving each cell a weight based on how close the cell's measured value is to the reference value gives the ResNet model more information on where a device is located thus giving a better location estimate compared to assigning a binary value to each cell.
Additionally, both variants of our localization method outperformed all four baselines. Essentially, the worst results of our localization method were better than the best performing baseline models.

Frequency fusion also provided some benefits as the propagation behaviour of multiple frequencies gave most methods additional information to make better location estimates.
Upstream fusion provided better location estimates compared to downstream fusion and both fusion methods beat the 5.0 GHz location estimates with the exception of the ResNet model that takes binary matrices as input.
We hypothesize that the small performance loss for the downstream fusion for this model compared to its 5.0 GHz counterpart is caused by some of the fused location estimates being worse than the best single frequency location estimate for some particular locations. 

Notably, our augmentation method outperforms the augmentation method proposed in extendGAN+.
Since our approach can generate thousands of realistic RSS samples from only a small number of calibration measurements, it can produce a training set that is an order of magnitude larger than that generated by the extendGAN+ augmentation method.
Moreover, our augmentation process is substantially faster: generating 23,705 samples takes approximately \~5.5 minutes, compared with approximately \~11 minutes to generate only about 600 samples using extendGAN+.

Seeding 20\% of the real-life data into the training set did not yield a meaningful improvement in localization performance for any of our methods or their fusion variants; in some cases the seeded regime is in fact slightly worse, suggesting that the synthetic dataset on its own already captures most of the signal that the methods can exploit.

\begin{table}[htbp]
\centering
\setlength{\tabcolsep}{4pt}
\caption{Comparison of localization errors across methods and data regimes (Mean $\pm$ Standard Deviation). Values that are \textbf{bold} or \underline{underline} indicate best and second best real localization error respectively for each data regimes on the test set.}
\label{tab:full_results}
\small
\begin{tabular}{lcccccccc}
\toprule
 & \multicolumn{4}{c}{\textbf{Synthetic Data Only}} 
 & \multicolumn{4}{c}{\textbf{Real-life Seeded Data}} \\
\cmidrule(lr){2-5} \cmidrule(lr){6-9}
\textbf{Localization Method} 
& \multicolumn{2}{c}{\textbf{Simulated Error (m)}} 
& \multicolumn{2}{c}{\textbf{Real Error (m)}}
& \multicolumn{2}{c}{\textbf{Simulated Error (m)}} 
& \multicolumn{2}{c}{\textbf{Real Error (m)}} \\
\cmidrule(lr){2-3} \cmidrule(lr){4-5}
\cmidrule(lr){6-7} \cmidrule(lr){8-9}
& 2.4\,GHz & 5.0\,GHz & 2.4\,GHz & 5.0\,GHz
& 2.4\,GHz & 5.0\,GHz & 2.4\,GHz & 5.0\,GHz \\
\midrule
\makecell[l]{ResNet w/ Binary Matrices}
& 0.52$\pm$0.50 & 0.61$\pm$0.67 & 3.86$\pm$2.18 & 4.30$\pm$2.60
& 0.51$\pm$0.54 & 0.61$\pm$0.66 & 3.74$\pm$2.06 & 4.15$\pm$2.38 \\

+ Downstream Fusion
& \multicolumn{2}{c}{0.60$\pm$0.60}
& \multicolumn{2}{c}{4.46$\pm$2.84}
& \multicolumn{2}{c}{0.59$\pm$0.61}
& \multicolumn{2}{c}{4.21$\pm$2.58} \\

+ Upstream Fusion
& \multicolumn{2}{c}{0.40$\pm$0.45}
& \multicolumn{2}{c}{3.78$\pm$2.22}
& \multicolumn{2}{c}{0.34$\pm$0.35}
& \multicolumn{2}{c}{3.65$\pm$2.04} \\

\midrule
\makecell[l]{ResNet w/ Multivalued Matrices}
& 0.22$\pm$0.16 & 0.83$\pm$1.28 & \underline{3.56$\pm$1.95} & 4.21$\pm$2.40
& 0.23$\pm$0.19 & 0.86$\pm$1.26 & \underline{3.55$\pm$1.87} & 4.43$\pm$2.92 \\

+ Downstream Fusion
& \multicolumn{2}{c}{0.56$\pm$1.05}
& \multicolumn{2}{c}{4.13$\pm$2.44}
& \multicolumn{2}{c}{0.59$\pm$0.99}
& \multicolumn{2}{c}{4.29$\pm$2.76} \\

+ Upstream Fusion
& \multicolumn{2}{c}{0.21$\pm$0.15}
& \multicolumn{2}{c}{\textbf{3.05$\pm$1.52}}
& \multicolumn{2}{c}{0.20$\pm$0.15}
& \multicolumn{2}{c}{\textbf{3.17$\pm$1.59}} \\

\midrule \midrule
Convex Hull
& 4.53$\pm$3.02 & 5.97$\pm$4.17 & 5.58$\pm$2.88 & 6.36$\pm$3.39
& 4.52$\pm$3.02 & 5.97$\pm$4.17 & 5.65$\pm$2.93 & 6.31$\pm$3.28 \\

\makecell[l]{+ Downstream Point Fusion}
& \multicolumn{2}{c}{5.03$\pm$3.39}
& \multicolumn{2}{c}{5.60$\pm$2.71}
& \multicolumn{2}{c}{5.03$\pm$3.39}
& \multicolumn{2}{c}{5.60$\pm$2.70} \\

+ Upstream Fusion
& \multicolumn{2}{c}{4.54$\pm$3.12}
& \multicolumn{2}{c}{5.13$\pm$2.61}
& \multicolumn{2}{c}{4.54$\pm$3.12}
& \multicolumn{2}{c}{5.04$\pm$2.56} \\

\midrule
DBSCAN
& 2.26$\pm$5.66 & 2.82$\pm$6.65 & 7.06$\pm$5.29 & 14.75$\pm$11.63
& 2.26$\pm$5.66 & 2.82$\pm$6.66 & 7.17$\pm$5.29 & 14.72$\pm$11.64 \\

\makecell[l]{+ Downstream Point Fusion}
& \multicolumn{2}{c}{2.33$\pm$4.42}
& \multicolumn{2}{c}{9.03$\pm$5.97}
& \multicolumn{2}{c}{2.34$\pm$4.42}
& \multicolumn{2}{c}{8.98$\pm$5.86} \\

\makecell[l]{+ Downstream Overlap Fusion}
& \multicolumn{2}{c}{2.17$\pm$4.43}
& \multicolumn{2}{c}{8.96$\pm$5.99}
& \multicolumn{2}{c}{2.17$\pm$4.44}
& \multicolumn{2}{c}{8.90$\pm$5.90} \\

+ Upstream Fusion
& \multicolumn{2}{c}{0.80$\pm$2.75}
& \multicolumn{2}{c}{9.85$\pm$9.80}
& \multicolumn{2}{c}{0.81$\pm$2.76}
& \multicolumn{2}{c}{9.69$\pm$9.68} \\

\midrule
RBF
& \multicolumn{2}{c}{-}
& \multicolumn{2}{c}{-}
& 4.52$\pm$2.92 & -
& 5.98$\pm$3.49 & - \\

\midrule

\makecell[l]{extendGAN+ with\\Ray-Traced Augmentation}
& \multicolumn{2}{c}{-}
& \multicolumn{2}{c}{-}
& 0.35$\pm$0.25 & -
& 4.59$\pm$2.59 & - \\

\midrule

extendGAN+
& \multicolumn{2}{c}{-}
& \multicolumn{2}{c}{-}
& 3.52$\pm$2.91 & -
& 6.87$\pm$3.13 & - \\
\bottomrule
\end{tabular}
\end{table}

\section{Conclusion and Future Work}
\label{sec:conclusions}

Indoor localization can benefit from advances in deep learning if the need for RSS fingerprints can be substantially reduced. As building information models become increasingly available for commercial and office buildings, there is an opportunity to leverage these models to simulate RF propagation and generate synthetic data for indoor localization.
In this paper, we show that a GPU-accelerated, ray tracing-based RF propagation tool such as Sionna RT can produce RSS estimates that are sufficiently close to their real-world counterparts to support the training of expressive deep learning models for indoor localization, provided that the underlying building model is properly calibrated. We further show that the building model can be calibrated using RSS fingerprints collected at a small number of locations; in our experiments, 34 locations on the first floor of a campus building were sufficient.
Our calibration procedure consists of two components: refinement of the building model's material parameters and a per-AP device calibration step that accounts for residual errors not captured by Sionna RT. With both calibration steps in place, the simulator achieves a mean absolute error of 4.97 dBm on the 2.4 GHz band and 6.47 dBm on the 5 GHz band relative to real measurements.

We evaluate the proposed localization method against four baselines spanning both learning-based and non-learning approaches. Our results show that transforming RSS values into multivalued matrices similar to grayscale images provides the best-performing representation and, when combined with upstream frequency fusion, enables our localization architecture to achieve a mean error of 3.05 meters on the real-data test set when trained exclusively on synthetic data. This outperforms the best-performing baseline, extendGAN+, which achieves a mean error of 4.59 meters using the same augmented dataset.

Two further observations are noteworthy. First, seeding the training set with 20\% of the real data does not yield a meaningful improvement over training exclusively on synthetic data, suggesting that the calibrated synthetic dataset already captures most of the information that the localization models can exploit. Second, frequency fusion of the 2.4 and 5 GHz bands consistently improves localization performance over either band alone for most methods, providing empirical evidence for the benefit of cross-band data fusion, even in its simplest form. Taken together, by combining calibrated ray-tracing-based RF modelling, a novel image-based representation of RSS values, frequency fusion, and deep learning, we enable accurate indoor localization using predominantly synthetic data.

In future work, we plan to extend the proposed calibration scheme to simultaneously support multiple receiving devices, including devices from different models and vendors, such that a single calibrated radio map can serve a heterogeneous user population without per-device retuning. We will also evaluate our localization method in multi-storey buildings, where the estimated location is represented as a triple $(x, y, l)$, with $l$ denoting the floor, as well as in more architecturally complex environments with non-rectangular rooms and a wider variety of structural materials. Finally, we aim to predict indoor movement trajectories by applying appropriate filtering techniques on top of the localization model, enabling applications such as collision avoidance between workers and automated guided vehicles (AGVs), and workflow optimization in warehouses and other industrial environments.

\begin{acks}
This research was supported in part by funding from the Natural Sciences and Engineering Research Council of Canada (RGPIN-2025-04316 and RGPIN-2021-03423),
Alberta Innovates and Western Economic Diversification Canada.
\end{acks}

\section*{Ethics and Privacy Statement}
Indoor localization can potentially expose sensitive information about the location and movement of individuals, creating risks of unauthorized tracking or surveillance. Our approach reduces the need to collect large amounts of real-world RSS fingerprint data by relying primarily on synthetic data, thereby limiting the collection of potentially sensitive location information. The passive collection of the RSS fingerprints and the processing outlined in our work can be performed by, and be under the complete control of, an individual, used for their own purposes.  However, deployment in real-world environments to serve many users, might involve localization computation and RSS fingerprint storage on platforms outside the control of the users. In such cases, appropriate consent, data governance, and safeguards against misuse should be incorporated. Additionally, it is necessary to ensure that localization errors do not create safety risks in applications where location information informs operational decisions.

\bibliographystyle{ACM-Reference-Format}
\bibliography{references_bibtex_zotero}

\end{document}